\documentclass[authoryear,preprint,12pt]{elsarticle}
\usepackage[a4paper,includehead,top=20mm,bottom=20mm,right=20mm,left=20mm]{geometry}
\usepackage{graphicx}
\usepackage{amsmath, amsthm, amsxtra}
\usepackage{array}
\usepackage{mathrsfs}
\DeclareMathAlphabet{\mathscr}{OT1}{pzc}{m}{it} 
\usepackage{amssymb}
\usepackage{stmaryrd}
\usepackage{eucal}
\usepackage{bm}
\usepackage{lscape}
\usepackage{xcolor}
\usepackage{setspace}
\usepackage{subfig}
\usepackage{upgreek}
\usepackage{listings}
\usepackage{nicefrac}
\usepackage{bbm}
\usepackage{lmodern}
\usepackage{nicefrac}
\usepackage{booktabs}
\usepackage{multirow}
\usepackage{algorithmic}
\usepackage{algorithm}
\usepackage[english]{babel}
\usepackage{hyperref}

\usepackage{hyphenat}
\newcommand{\Dt}[1]{#1^{\scalebox{0.4}{\textbullet} } }

\renewcommand{\t}[1]{\bm{#1}}
\renewcommand{\tt}[1]{\pmb{#1}}
\newcommand{\p}{\partial}
\renewcommand{\d}{\, \mathrm d }
\newcommand{\dd}[2]{\frac{\d #1}{\d #2}}
\newcommand{\pd}[2]{\frac{\p #1}{\p #2}}
\renewcommand{\comma}{\ , \quad }
\newcommand{\Om}{\Omega}

\newcommand{\curl}{\operatorname{curl} }

\newcommand{\del}{\updelta}
\newcommand{\Del}{\Delta}

\newcommand{\Bo}{{\B_0}}
\renewcommand{\S}{\mathcal S}
\newcommand{\eps}{\varepsilon}

\newcommand{\Ref}{\text{\tiny ref.}}
\newcommand{\fr}{\text{fr.}}
\newcommand{\bo}{\text{bo.}}

\newcommand{\begeq}{\begin{equation}\begin{gathered}}
\newcommand{\eqend}{\end{gathered}\end{equation}}
\newcommand{\el}{\text{\tiny el.}}
\newcommand{\ma}{\text{\tiny mag.}}
\newcommand{\ele}{\text{\tiny e}}
\newcommand{\Body}{\mathscr b}
\newcommand{\B}{\mathcal B}

\newcommand{\EE}{\mathscr E}
\newcommand{\BB}{\mathscr B}

\newcommand{\MM}{\mathscr M}
\newcommand{\mm}{\mathscr m}
\newcommand{\HH}{\mathscr H}
\newcommand{\JJ}{\mathscr J}
\newcommand{\DD}{\mathscr D}
\newcommand{\mH}{\mathfrak H}
\newcommand{\mD}{\mathfrak D}

\newcommand{\elmo}{\mathscr G}
\newcommand{\elF}{\mathscr F}

\newcommand{\field}{\text{\tiny f.}}
\newcommand{\matter}{\text{\tiny m.}}
\newcommand{\free}{\mathscr f}

\newcommand{\jl}{\llbracket}
\newcommand{\jr}{\rrbracket}
\newcommand{\Po}{\mathscr P}
\renewcommand{\v}{\mathrm v}
\newcommand{\pre}{\mathscr p}

\newcommand{\Bs}{B_\text{s}}

\journal{Journal of Computational Physics}

\begin{document}

\begin{frontmatter}

\title{Theory and computation of electromagnetic fields and thermomechanical structure interaction for systems undergoing large deformations \vspace{5mm}\\ 
\small The post-print version of the manuscript: \\
B.E. Abali, A.F. Queiruga, J. Comput. Phys. (2019), pp. 1--32 \\ https://doi.org/10.1016/j.jcp.2019.05.045
}
\author{B. E. Abali\corref{cor1}}
\ead{bilenemek@abali.org}
\address{Technische Universit\"at Berlin, Germany}
\cortext[cor1]{Corresponding author}
\author{A. F. Queiruga}
\ead{afqueiruga@lbl.gov}
\address{Lawrence Berkeley National Laboratory, USA}

\begin{abstract}
For an accurate description of electromagneto\hyp{}thermomechanical systems, electromagnetic fields need to be described in a \textsc{Euler}ian frame, whereby the thermomechanics is solved in a \textsc{Lagrange}an frame. It is possible to map the \textsc{Euler}ian frame to the current placement of the matter and the \textsc{Lagrange}an frame to a reference placement. We present a rigorous and thermodynamically consistent derivation of governing equations for fully coupled electromagneto\hyp{}thermomechanical systems properly handling finite deformations. A clear separation of the different frames is necessary. There are various attempts to formulate electromagnetism in the \textsc{Lagrange}an frame, or even to compute all fields in the current placement. Both formulations are challenging and heavily discussed in the literature. In this work, we propose another solution scheme that exploits the capabilities of advanced computational tools. Instead of amending the formulation, we can solve thermomechanics in the \textsc{Lagrange}an frame and electromagnetism in the \textsc{Euler}ian frame and manage the interaction between the fields. The approach is similar to its analog in fluid structure interaction, but more challenging because the field equations in electromagnetism must also be solved within the solid body while following their own different set of transformation rules. We additionally present a mesh-morphing algorithm necessary to accommodate finite deformations to solve the electromagnetic fields outside of the material body. We illustrate the use of the new formulation by developing an open-source implementation using the FEniCS package and applying this implementation to several engineering problems in electromagnetic structure interaction undergoing large deformations.
\end{abstract}

\end{frontmatter}
\section{Introduction}

The theory of electromagnetism started with \cite{maxwell1892} and is often explained by \textsc{Maxwell}'s equations. The theory has been continuously developed and amended, notably in the 1950s during the so-called renaissance of thermodynamics. The inclusion of mechanics and thermodynamics into the theory of electromagnetism can be modeled by using balance equations; however, there is no consensus about the correct form of the balance equations among the scientific community. The lack of consensus owes to various challenges in the formulation and the lack of experimental verifications for proposed formulations. For example, there are different representations of \textsc{Maxwell}'s equations, cf.\ \cite{pao1975} and \cite[Sect.\,II]{chu1966}. Another challenge occurs due to the different invariance properties of balance laws and \textsc{Maxwell}'s equations, raising questions about the proper forms of electromagnetic interaction equations in matter. The readers are directed to \cite[\textsection 286]{truesdell_toupin} for some of these different formulations. \\

In addition to agreeing upon the balance equations for electromagneto\hyp{}thermomechanical fields, we must also define the constitutive responses, i.e., the equations modeling the material behavior. Typically, phenomenological equations are constructed relying on experiments, which limit their applicability to the particular conditions of the measurement conditions. In order to define generic relations, we want to follow a consistent theoretical derivation through thermodynamics. However, there is no consensus for deriving thermodynamically sound constitutive equations for electromagnetically polarizable systems. The challenge again lies in the formulation of balance equations, especially on the balance of energy, which has been discussed by \cite{ericksen2007} as well as in \cite{steigmann2009formulation}. There exist a few complete theories for polarized deformable media, as those of \cite[Chap.\,XIII]{groot1984}, \cite[Chap.\,9]{mueller1985}, \cite[Chap.\,5]{eringen1990}, \cite[Chap.\,15]{kovetz2000}, \cite{brechet2014}, and \cite[Chap.\,3]{027}. Each of the mentioned formulations is different, and an experimental verification to determine their correctness is still missing. \\

Computational methods help us to simulate and comprehend realistic applications in two ways. First, we can estimate the response of a system before manufacturing. Secondly, we can design experiments for validating or even discovering an accurate representation of the physical world. Several computational strategies exist for solving coupled equations by means of finite element simulations. For detailed reviews, see \cite{benjeddou2000}, \cite{hachkevych2004}, and \cite{vidal2011}. Different simplifications of the governing equations are employed in order to enable a numerical analysis. Especially solving coupled problems involves numerical challenges and different numerical treatments are proposed for solving coupled problems, see \cite{chung2014three, jin2015finite, gil2016new, dorfmann2017nonlinear, assous2017mathematical, demkowicz2017finite, pierrus2018solved}. Different length and time scales are incorporated for a possible modeling in \cite{zah2015multiplicative, schroeder2016algorithmic, zhang2017reduced, keip2018multiscale}. Combining different numerical techniques also yields feasible methods in computational modeling, see \cite{liu2016solutions, liu2016smoothed, nedjar2017coupled, kraus2017gempic, lanteri2018multiscale, kodjo2019multiscale}. General formulations for thin structures are studied in \cite{klinkel2013solid, staudigl2018nonlinear, chroscielewski2018nonlinear}. For example, restriction to the quasi-static case by neglecting inertial terms can be seen in \cite{yi1999}, \cite{ahmad2006} and \cite{queiruga2016microscale}. A case without free charges was presented by \cite{svendsen2005}. A magneto-elasto-static case has been suggested in \cite{spieler2014}, \cite{glane2017}. In \cite{mehnert2017} the temperature distribution is also computed by neglecting inertial terms. A complete dynamical description and transient computation of electromagneto\hyp{}thermomechanical has been proposed in \cite{queiruga2016} and \cite{abali2017}. In most of these works, the formulations are established on the same configuration. If the electromagnetic fields interact with solid bodies, a \textsc{Lagrange}an frame is chosen, where each coordinate maps to a material particle. In the case of fluids, a \textsc{Euler}ian frame is chosen, wherein each coordinate indicates a fixed position in ordinary (physical) space. Solving electromagnetic fields in a \textsc{Euler}ian frame and thermomechanical fields in a \textsc{Lagrange}an frame is not a new idea. Among others, \cite{kankanala2004, rieben2007, stiemer2009, barham2010, steinmann2011computational, skatulla2012multiplicative, vogel2013, ethiraj2016multiplicative, pelteret2016} have developed computational strategies to overcome different problems. In all aforementioned works, governing equations differ due to the different simplifications and assumptions used. Instead of a comparison of different works, we start from the beginning with a new derivation of the equations based on continuum mechanics such that any assumptions and weaknesses in the methodology can be precisely identified and addressed.  \\

We begin by outlining the theory in Section \ref{sec:governing}, following \cite[Chap.\,3]{027} most closely. The main objective is to compute the \textit{primitive} variables for solids under finite deformation, namely the temperature $T$ and displacement $\t u$, and to compute for the entirety of space encompassed by the computational domain the so-called electromagnetic potentials $\phi$, $\t A$. In the formulation we will use different frames, where $\t X$ denotes the reference position of a massive particle, and $\t x$ indicates a position in the ordinary space. The formulation referring to the placement of particles in $\t X$ is called the \textsc{Lagrange}an frame (placement, configuration). Thermomechanical fields belong to massive particles such that they are computed in the \textsc{Lagrange}an frame, which allows to incorporate large deformations for a material system. Electomagnetic fields propagate in $\t x$ with or without interacting with material such that their formulation is developed in the \textsc{Euler}ian frame (configuration), which is tantamount to the control volume for an open system. In order to close the formulation, we develop thermodynamically consistent constitutive equations for solids in Section \ref{sec:constitutive}. The theory is limited to elastic materials; plasticity is not treated. The constitutive equations are developed for polarized materials such that all coupling effects, including piezo-and pyroelectric and thermal expansion, are captured precisely. Therefore, the formulation gives rise to coupled and nonlinear field equations to be solved. We discuss the issues that arise when solving these equations using the finite element method in Section \ref{sec:computational}. In order to address the large deformations of the mesh of the solid body embedded in the mesh of the electromagnetic computational domain, we present a mesh morphing algorithm that enables the calculations by keeping a valid mesh in the space surrounding the body. The variational forms and new algorithms are implemented with the aid of the novel collection of open-source packages provided by the FEniCS project \citep{fenics, fenics_book}. The library containing the presented mesh morphing algorithm and other helper routines is released at \url{https://github.com/afqueiruga/afqsfenicsutil} under the GNU Lesser General Public License \citep{gnupublic}. In Section \ref{sec:examples}, we present three simulations of example applications to electromagnetic devices. The simulation setups and FEM implementations of the variational forms are published at \url{https://github.com/afqueiruga/EMSI-2018} under the GNU General Public License. We conclude the discussion in Section \ref{sec:conclusion}. 

\section{Governing equations}
\label{sec:governing}
Consider a solid body $\Body$ immersed in air. We will solve electromagnetic fields in the whole domain $\Om$ including a body $\Body$ and air, $\Omega_\textrm{air}=\Om\backslash \Body$. The solid body undergoes a deformation. The mechanical fields will be computed within the body $\Body$. Although the surrounding air might be set in motion due to the deformation of the solid body and its own electromagnetic interactions, we will ignore the fluid motion. For certain applications, we might want to determine the temperature distribution within the air as well, but it is not of interest for now. We choose to compute the temperature distribution only within the solid body to save computational time. Therefore, we aim at determining governing equations for electromagnetic fields within $\Om$ and for thermomechanical fields within $\Body$. At the interface $I=\Om\cap\p\Body$, we need to discuss the interaction and model by satisfying an additional set of equations derived in a rational approach. We motivate the theory in three subsections: 
\begin{itemize}
\item specifying the partial differential equations modeling electromagnetic fields in the whole domain; 
\item specifying the partial differential equations describing thermomechanical fields in the solid body; and
\item specifying the jump conditions on the interface between the solid body and its surroundings.
\end{itemize}
We will use Cartesian coordinates and the usual tensor index notation with the \textsc{Einstein} summation convention over repeated indices. Note that different typefaces will be used to denote the electromagnetic fields measured in different frames.

\subsection{Electromagnetic fields}

The main objective in electromagnetism is to obtain the electric field, $\t E$, and the magnetic flux density, $\t B$ (an area density). SI units are the most appropriate choice for thermomechanical couplings, where the electric field is measured in V(olt)/m(eter) and the magnetic flux density is measured in T(esla). We start off with \textsc{Faraday}'s law:
\begeq \label{global.faraday}
\Dt{\bigg( \int_S \BB_i \d a_i \bigg)} = - \int_{\p\S} \EE_i \d\ell_i \ ,
\eqend
defined on an arbitrarily moving surface $S$ with the electric field measured on the co-moving frame, $\tt\EE$, as well as the magnetic flux on the co-moving frame, $\tt\BB$. In other words, the measurement device is installed on $S$ and moves with it. The notation $\Dt{\left(\right)}$ denotes rate regarding the motion of the surface $S$. Assume that the domain $S$ defined in $\t x$ moves with the velocity $\Dt{\t x} = \t w$ measured with respect to the laboratory frame that is set to be fixed (not moving). In order to define a velocity as a measurable quantity, we have to declare one frame without possessing a velocity. Of course, a laboratory frame on Earth moves with respect to other planets and stars; however, we declare and maintain the laboratory frame as being fixed such that every motion detected in that frame acquires a well-defined velocity. Since $\tt\BB$ and $\tt\EE$ are detected on a moving frame, we need their transformations to the laboratory frame,
\begeq
\EE_i = E_i + (\t w \times \t B)_i \comma
\BB_i = B_i \ ,
\eqend
for the non-relativistic case, where the magnitude of the domain velocity is small with respect to the speed of light in vacuum, $|\t w|\ll c$. By using \textsc{Stokes}'s theorem as well as the identity for the derivative of a differential area
\begeq
\Dt{(\d a_i)} = \Big( \pd{w_k}{x_k} \delta_{ji} - \pd{w_j}{x_i} \Big) \d a_j \ ,
\eqend
we acquire the local form of \textsc{Faraday}'s law:
\begeq
\Dt B_j + B_i \Big( \pd{w_k}{x_k} \delta_{ji} - \pd{w_j}{x_i} \Big) = - \curl(\t E + \t w \times \t B)_j \ , \\
\pd{B_j}{t} + \pd{B_j}{x_k} w_k + B_j \pd{w_k}{x_k} - B_i \pd{w_j}{x_i} + \epsilon_{jkl} \pd{E_l}{x_k} + \epsilon_{jkl}\epsilon_{lmn} \pd{w_m B_n}{x_k} = 0 \ , \\
\pd{B_j}{t} + \pd{B_j w_k}{x_k} - B_i \pd{w_j}{x_i} + \epsilon_{jkl} \pd{E_l}{x_k} + \pd{w_j B_k}{x_k} - \pd{w_k B_j}{x_k}= 0 \ , \\
\pd{B_j}{t}  + \epsilon_{jkl} \pd{E_l}{x_k} + w_j \pd{B_k}{x_k} = 0 \ , 
\eqend
using the identity  $\epsilon_{jkl}\epsilon_{lmn} = \delta_{jm}\delta_{kn}-\delta_{jn}\delta_{km}$ with the \textsc{Kronecker} delta, $\delta_{ij}$, and the \textsc{Levi-Civita} symbol, $\epsilon_{ijk}$. Moreover, we can consider the special case where the surface $S$ is a closed hull, for example the boundary of a continuum body, $\p\Body$, without a line boundary such that the right-hand side in Eq.\,\eqref{global.faraday} vanishes and we obtain after an integration in time
\begeq
\int_{\p\Body} B_i \d a_i = \text{const.}|_t \ .
\eqend
If we select the initial magnetic flux as zero, the integration constant drops. Since the selected boundary is a closed hull, we can apply \textsc{Gauss}'s law and acquire
\begeq
\pd{B_i}{x_i} = 0 \ .
\eqend
We have obtained the so-called first set of \textsc{Maxwell}'s equations:
\begeq
\pd{B_i}{x_i} = 0 \comma 
\pd{B_i}{t} + \epsilon_{ijk} \pd{E_k}{x_j} = 0 \ .
\eqend
These equations are universal; i.e., they hold for any material and even in the case of no massive particles (vacuum). Hence, the coordinate $\t x$ denotes a location or point in the (ordinary) space. We call it a \textit{spatial} frame since the coordinates indicate a position in space. There might be a massive particle occupying the location $\t x$, but the coordinate still indicates a location in space without any relation to that particle or its motion. The sought-after electromagnetic fields, $\t E$ and $\t B$, have to satisfy the latter equations. Their solution is obtained by using the following \textit{ansatz} functions:
\begeq \label{em.ansatz}
E_i = - \pd{\phi}{x_i} - \pd{A_i}{t} \comma
B_i = \epsilon_{ijk} \pd{A_k}{x_j} \ ,
\eqend
such that now we search for the electric potential $\phi$ in V and magnetic potential $\t A$ in T\,m for $\forall \t x \in \Om$. If we can compute the electromagnetic potentials, we readily obtain the electromagnetic fields from the latter equations. Since we aim to describe the system using only four components $\{\phi, A_1, A_2, A_3\}$ instead of six components $\{E_1, E_2, E_3, B_1, B_2, B_3\}$, there are two scalar degrees of freedom that are not uniquely determined; namely $\p\phi/\p t$ and $\p A_i/\p x_i$ can be chosen freely. This so-called \textit{gauge} freedom can be used to eliminate many numerical problems \citep{baumanns2013}. We will use \textsc{Lorenz}'s gauge:
\begeq \label{lorenz}
\pd{\phi}{t} = -c^2 \pd{A_i}{x_i} \comma c^2 = \frac1{\mu_0\eps_0} \ , 
\eqend
with the speed of light in vacuum, $c$, defined by the precisely known universal constants:
\begeq
\eps_0 = 8.85\cdot 10^{-12} \,\text{A\,s/(V\,m)} \comma
\mu_0 = 12.6\cdot 10^{-7} \,\text{V\,s/(A\,m)} \ .
\eqend
In order to motivate the second set of \textsc{Maxwell}'s equations, we use the balance of electric charge in an open system with the control volume or domain $\Om$ where the domain moves with velocity $\t w$
\begeq
\Dt{\bigg( \int_\Om q \d v \bigg)} = \int_{\p\Om} \Big( q(w_i-v_i) - \JJ_i \Big) \d a_i \ .
\eqend
The electric charge density, $q$ in C(oulomb)/m$^3$, can be determined if we have a constitutive equation for the electric current (area density), $\tt\JJ$ in A(mpere)/m$^2$. If a massive particle conveying an electric charge of $q$ enters the domain, the amount of charge within the domain increases. The particle moves with $\t v$ and can enter the domain only across its boundary $\p\Om$. The particle is entering if the relative velocity $\t w-\t v$ is positive along the surface direction, $\d a_i=n_i \d a$ and exiting if it is in the other direction. We refer to \cite{mueller1983a}, \cite{mueller1983b} for a discussion of balance equations in an open system with a moving domain. The surface direction $\t n$ points outward from the domain. We can again get the local form after using the rate of the volume element in the spatial frame moving with $\t w$,
\begeq
\Dt{\d v} = \pd{w_i}{x_i} \d v \ ,
\eqend
and apply \textsc{Gauss}'s law,
\begeq
\Dt q + q \pd{w_i}{x_i}  = \pd{}{x_i}\Big( q(w_i-v_i) - \JJ_i \Big)  \ , \\
\pd{q}{t} + \pd{q w_i}{x_i} = \pd{}{x_i}\Big( q(w_i-v_i) - \JJ_i \Big)  \ , \\
\pd{q}{t} + \pd{J_i}{x_i} = 0 \ ,
\eqend
where $J_i = \JJ_i + q v_i$ represents the electric current measured in the laboratory frame. Since the domain $\Om$ has a well-defined boundary, $\p\Om$, we can introduce a charge potential $\tt\DD$ measured on the moving domain as follows
\begeq
\int_\Om q \d v = \int_{\p\Om} \DD_i \d a_i \ ,
\eqend
leading to the following \textsc{Maxwell} equation after applying \textsc{Gauss}'s law
\begeq
q = \pd{\DD_i}{x_i} \ .
\eqend
(Note the typeface on the quantity $\DD$.) The charge potential $\tt\DD$ in C/m$^2$ is quite general and stemming from the \textit{total} charge $q$ in space. The charge potential is also called dielectric displacement or electrical flux density in the literature. We emphasize that $\tt\DD$ is the total charge potential incorporating bound and free charges. Now by using the charge potential, we rewrite the balance of charge in an open domain with a closed boundary, $\p\p\Om=\{\}$,
\begeq
\Dt{\bigg( \int_{\p\Om} \DD_i \d a_i \bigg)} = \int_{\p\Om} \Big( q(w_i-v_i) - \JJ_i \Big) \d a_i \ ,
\eqend
as a balance equation on a surface with its boundary $\p S$
\begeq
\Dt{\bigg( \int_{S} \DD_i \d a_i \bigg)} = \int_{\p S} \HH_i \d\ell_i + \int_{S} \Big( q(w_i-v_i) - \JJ_i \Big) \d a_i \ ,
\eqend
where the flux on the surface boundary $\tt\HH$ is called the current potential (or also magnetic field strength). It is measured on the moving surface. Transformations of the charge and current potentials from the moving frame to the laboratory (fixed) frame read
\begeq
\DD_i = D_i \comma
\HH_i = H_i + (\t D\times \t w)_i \ ,
\eqend
for the non-relativistic case. Now we can insert the rate of the area element, apply \textsc{Stokes}'s theorem, and obtain the local form
\begeq
\Dt D_j + D_i \Big( \pd{w_k}{x_k} \delta_{ji} - \pd{w_j}{x_i}\Big) = \curl(\t H + \t D\times \t w)_j + q w_j - J_j \ , \\
\pd{D_j}{t} + \pd{D_j}{x_i} w_i + D_j \pd{w_k}{x_k} - D_i \pd{w_j}{x_i} = \epsilon_{jkl} \pd{ H_l}{x_k} + \epsilon_{jkl}\epsilon_{lmn} \pd{D_m w_n}{x_k} + q w_j - J_j \ , \\
\pd{D_j}{t} + \pd{D_j w_i}{x_i}  - D_i \pd{w_j}{x_i} = \epsilon_{jkl} \pd{ H_l}{x_k} + \pd{D_j w_k}{x_k} - \pd{D_k w_j}{x_k} + \pd{D_i}{x_i} w_j - J_j \ , \\
\pd{D_j}{t} = \epsilon_{jkl} \pd{ H_l}{x_k} - J_j \ .
\eqend
We have obtained the second set of \textsc{Maxwell}'s equations:
\begeq
q = \pd{D_i}{x_i} \comma
\pd{D_j}{t} = \epsilon_{jkl} \pd{H_l}{x_k} - J_j \ ,
\eqend
which are universal and hold in the whole domain. The measured charge and current potentials on the laboratory frame---$\t D$ and $\t H$ respectively---do not change with respect to a domain velocity $\t w$. Therefore, the domain velocity is arbitrary giving us a freedom to choose the domain velocity to our advantage. Later in the text, we will discuss a method to generate the domain velocity in such a way that the mesh quality remains optimal. \\

We have reached the following governing equations for the total electric charge, current, and their potentials: 
\begeq\label{governing.em}
\pd{q}{t} + \pd{J_i}{x_i} = 0 \comma
J_i = \JJ_i + q v_i \comma
q = \pd{D_i}{x_i} \ , \\
\pd{D_j}{t} - \epsilon_{jki} \pd{H_i}{x_k} +  J_j = 0 \ .
\eqend
The equations need to be used to compute the electromagnetic potentials, $\phi$ and $\t A$. In order to close the equations, the total charge potential and the total current potential need to be expressed in terms of the electromagnetic potentials. The \textsc{Maxwell--Lorentz} aether relations
\begeq \label{em.MLrelations}
D_i = \eps_0 E_i \comma H_i = \frac1{\mu_0} B_i \ ,
\eqend
augmented by Eq.\,\eqref{em.ansatz} presents the relation closing the coupled governing equations. These equations will be solved in the whole domain, $\Om$.

\subsection{Thermomechanical fields}

Consider a continuum body, $\Body$, within the domain $\Om$. This body consists of massive particles with electric charge. Mass (volume) density $\rho$ and specific charge (per mass) $z$ are material dependent variables. Their initial values are known. The total specific charge $z$ in a material is decomposed as \textit{free} charge $z^\fr$ and \textit{bound} charges $z^\bo$ as follows
\begeq
z = z^\fr + z^\bo \ .
\eqend
Free charges are the valence electrons carrying the electric current effectively in a conductor, they can move large distances. Bound charges are held by the intra-molecular forces and they only move less than the molecular length. Their motions give rise to a decomposition of the charge and current potentials,
\begeq \label{em.DHrelations}
D_i = \mD_i - P_i \comma
H_i = \mH_i + \MM_i \ ,
\eqend
where the bound charge potential $P_i$ is called an electric polarization and the bound current potential $\MM_i$ is called a magnetic polarization. The minus sign is a convention of the declaration of electric polarization in the atomistic scale. Since we already have introduced the \textsc{Maxwell--Lorentz} aether relation, we need constitutive equations either for $\mD_i$ and $\mH_i$ or for $P_i$ and $\MM_i$. By using the above definitions we achieve the analogous decomposition for the electric current:
\begeq
J_i = J_i^\fr + J_i^\bo \comma J_i^\bo = \pd{P_i}{t} + \epsilon_{ijk} \pd{\MM_k}{x_j} \ .
\eqend
See \ref{app.el.current.def} for its well-known derivation. \\

The massive particles' initial positions are known and denoted by $\t X$. Effected by mechanical, thermal, and electromagnetic forces, particles at $\t X$ displace as much as $\t u$ and move to $\t x$ such that $\t u = \t x - \t X$ in m. Moreover, it is necessary to compute the temperature $T$ in K(elvin) and electromagnetic potentials of particles. Since we know the initial positions of the (non-congruent) particles, we can use $\t X$ in order to identify the material particles. This configuration uses coordinates indicating material particles' positions at the reference placement. The reference placement is defined by the vanishing energy, which will be investigated in the next section using thermodynamics. Initially, we start from the reference placement and the amount of particles remains the same. This configuration is called a material system expressed in the \textsc{Lagrange}an frame with $\t X$ denoting the initial placement of particles. In the \textsc{Lagrange}an frame, we search for $\t u$, $T$, as functions in space $\t X$ and time $t$. Their field equations are given by the balance equations at the current placement. As the space, $\t X$, indicates the same particle throughout the simulation, we can introduce the balance equations at the current placement and a transformation between current and initial placement. Therefore, in a material system, we start with balance equations for mass, linear momentum, and energy at the current placement and then determine the field equations in the \textsc{Lagrange}an frame by transformation into the current placement. The equations are finally closed by the constitutive equations. We start off with the general balance equation in a volume
\begeq \label{general.balance}
\Dt{\bigg( \int_\Body \psi \d v \bigg)} 
= \int_{\p\Body} \Phi_j \d a_j + \int_\Body k \d v + \int_\Body p \d v \comma
\eqend
where rate of the volume density $\psi$ is balanced by the fluxes across the boundary $\Phi_j$, volumetric supply terms $k$, and production terms $p$. Mathematically, supply and production terms are identical; however, we handle them separately as we can control supply terms but fail to steer production terms. The general balance equation in Eq.\,\eqref{general.balance} is defined at the current placement, where $\t x$ denotes the current positions of material particles conforming a material system. In other words, the integral measure moves with material particles such that the integration domain is the current placement of the continuum body. We start a simulation with known initial conditions, particles at $\t X$, and compute their motion to $\t x$. The current positions of particles change in time such that $\Dt x_i = v_i$. The rate is defined with respect to the material particle. Now, the rate of the infinitesimal volume element (an integral measure) reads
\begeq
\Dt{\d v} = \pd{v_i}{x_i} \d v \ ,
\eqend
leading to the following local form after applying \textsc{Gauss}'s law:
 \begeq
 \pd{\psi}{t} + \pd{}{x_j} ( v_j \psi - \Phi_j ) - k = p \ .
 \eqend
In the local form we write the production term on the right-hand side for a clear separation of conserved quantities. If the production term vanishes, the variable in the balance equation is a conserved quantity. We axiomatically start with the balance equations for the mass, total momentum, and total energy as given in Table~\ref{tab:supply.flux}.
\begin{table}[!hbt]
\centering
\caption{Matter and (electromagnetic) field related volume densities in the balance equations and their supply terms, flux terms, and production terms.}
\renewcommand{\arraystretch}{1.3} 
\begin{tabular}{c|c|c|c}
\toprule
$\psi$ &
$\Phi_j$ &
$k$ & 
$p$ \\
\midrule
$\rho$ & $0$ & $0$ & $0$ \\
$\rho v_i + \elmo_i$ & $v_j \elmo_i + \sigma_{ji} + m_{ji}$ & $\rho f_i$ & $0$ \\
$\rho e^\matter + e^\field$ & $v_j e^\field + \zeta_j + \Po_j$ & $\rho s$  & $0$\\
\bottomrule
\end{tabular}
\label{tab:supply.flux}
\end{table}
It is important to emphasize that we assume that mass, total momentum, and total energy are all conserved quantities. We skip a long discussion about the angular momentum and simply assume that the material is non-polar, leading to a vanishing spin density such that the angular momentum reduces to the moment of momentum. In this case, the balance of angular momentum is fulfilled by having a symmetric non-convective flux of linear momentum, $\t\sigma+\t m$. After using Table~\ref{tab:supply.flux}, the balance equations read
\begeq
\pd{\rho}{t} + \pd{v_j \rho}{x_j} = 0 \ , \\
\pd{}{t}(\rho v_i + \elmo_i) + \pd{}{x_j} \big( v_j \rho v_i - \sigma_{ji} - m_{ji}\big) - \rho f_i = 0 \ , \\
\pd{}{t}(\rho e^\matter + e^\field) + \pd{}{x_j} \big( v_j \rho e^\matter - \zeta_j - \Po_j\big) - \rho s = 0 \ 
\eqend
Mass density, $\rho$, has a convective flux, $\t v \rho$, because mass is conveyed by the moving material particles. Total momentum density, $\rho \t v + \tt\elmo$, consists of a part due to matter, $\rho \t v$, and another part due to the electromagnetic field, $\tt\elmo$. Matter and the electromagnetic field are coupled; however, we will be decomposing terms by splitting the fields along their interaction. Consider a massive object moving in an electromagnetic field in a way that the electromagnetic field does not alter, i.e., matter and field are independent. Of course, as given in the balance of momentum, the existing field's rate applies forces on the moving charges and a massive object has usually (bound) electric charges such that its acceleration leads to a change in the velocity, in other words, matter and field are coupled but they are independent. As they are independent, we treat the electromagnetic field and matter separately (independently) in a coupled manner. We can always fix matter and vary the field, and \textit{vice versa}. \\

In the balance of momentum, convective flux affects terms related to matter but not field. Non-convective flux of momentum, $\t\sigma + \t m$, is also decomposed into \textsc{Cauchy}'s stress, $\t \sigma$, and an electromagnetic stress, $\t m$. The specific supply term, $\t f$, is the (known) body force because of gravity. Total energy density, $\rho e^\matter+e^\field$, is decomposed into matter and field energies, as well as non-convective fluxes, $\tt\zeta$ and $\tt\Po$, respectively. Again, only the energy due to matter is conveyed by moving massive particles, $\t v \rho e^\matter$, as a convective flux. The specific supply term, $s$, is considered as given. All the other terms will be defined in the following discussion. \\

In the above formulation, the electromagnetic momentum, stress, energy, and flux are the key terms for the correct interaction. Hence it is customary to introduce the following relations:
\begeq \label{el.mag.momentum.energy}
\pd{\elmo_i}{t} = \pd{m_{ji}}{x_j} - \elF_i \ , \\
\pd{e^\field}{t} = \pd{\Po_j}{x_j} - \pi \ ,
\eqend
where the electromagnetic momentum $\tt\elmo$ and electromagnetic stress $\t m$ are related to the electromagnetic force (density), $\tt\elF$; analogously, electromagnetic energy (density) $e^\field$ and electromagnetic flux $\tt\Po$ are related to an electromagnetic power (density), $\pi$. These mathematical identities might be called balance equations; however, we refrain ourselves from using this terminology, since there is an ongoing discussion in the literature about the correctness of this terminology. It is obvious that we can insert the latter identities and renew the table as in Table~\ref{tab:supply.flux.2}. \\
\begin{table}[!hbt]
\centering
\caption{Matter related volume densities in the balance equations and their supply terms, flux terms, and production terms.} 
\renewcommand{\arraystretch}{1.3}
\begin{tabular}{c|c|c|c}
\toprule
$\psi$ &
$\Phi_j$ &
$k$ & 
$p$ \\
\midrule 
$\rho$ & $0$ & $0$ & $0$ \\
$\rho v_i$ & $\sigma_{ji}$ & $\rho f_i$ & $\elF_i$ \\
$\rho e^\matter$ & $\zeta_j$ & $\rho s$  & $\pi$\\
\bottomrule
\end{tabular}
\label{tab:supply.flux.2}
\end{table}

We stress these balance equations belong to the quantities related to matter; for the momentum (of matter) and energy (of matter), they read
\begeq
\pd{\rho v_i}{t} + \pd{}{x_j} \big( v_j \rho v_i - \sigma_{ji} \big) - \rho f_i = \elF_i \ , \\
\pd{\rho e^\matter}{t} + \pd{}{x_j} \big( v_j \rho e^\matter - \zeta_j \big) - \rho s = \pi \ .
\eqend
After using the balance of mass and the material derivative 
\begeq
\dd{}{t}= \pd{}{t} + v_i\pd{}{x_i} \ ,
\eqend
they are
\begeq
\rho \dd{ v_i}{t} - \pd{\sigma_{ji} }{x_j} - \rho f_i = \elF_i \ , \\
\rho \dd{ e^\matter}{t} - \pd{\zeta_i}{x_i} - \rho s = \pi \ ,
\eqend
furnishing the consequence that momentum and energy of matter are not conserved quantities in the case of electromagnetism. \\

The production terms $\tt\elF$ and $\pi$ need to be defined in such a way that they vanish if electromagnetic fields are zero. Unfortunately, their definitions are challenging and there exists no consensus in the scientific community; see for example \cite{obukhov2008, mansuripur2010, griffiths2012, bethune2015}. We will propose terms in accordance with Eq.\,\eqref{el.mag.momentum.energy}, which is the method of derivation used in \cite[Eq.\,(15)]{lorentz1904}, \cite[Chap.\,1]{jones1964}, \cite[Chap.\,XIV]{groot1984}, \cite[Chap.\,8]{griffiths1999}, \cite[Sect.\,3.3]{low2004}. If the electromagnetic momentum, $\elmo_i$, is defined, then, as a consequence of Eq.\,\eqref{el.mag.momentum.energy}$_1$, we can deduce the electromagnetic stress, $m_{ji}$, and the electromagnetic force density, $\elF_i$. By following \cite{barnett2010} we emphasize that different choices are perfectly appropriate. The reasoning can be explained as followed: the manifestation of a force as a contact force leads a term into the electromagnetic stress, whereas as a body force leads to a term causing a momentum rate. In the atomistic scale we know that all electromagnetic forces are contact forces. However, in the macroscopic scale we can observe a momentum change due to the electromagnetic fields such that declaring a body force is also suitable. Any choice of $\tt\elmo$, $\t m$, and $\tt\elF$ is possible as long as the relations in Eq.\,\eqref{el.mag.momentum.energy} are fulfilled. Analogously, we can choose an electromagnetic flux, $\tt\Po$, leading to the field energy and power. The choices cannot be justified or falsified by experiments, since we cannot detect contact forces and motion independently. Every sensor---used for detecting contact forces---depends on material properties coupling motion with electromagnetism. \\

Now we introduce a specific (per mass) internal energy, $u$, by decomposing the energy of matter in kinetic and internal energy
\begeq
e^\matter = \frac12 v_i v_i + u \ .
\eqend
By inserting the latter into the balance of energy and using the balance of momentum, 
\begeq
\rho \dd{v_i}{t} v_i + \rho \dd{u}{t} - \pd{\zeta_j}{x_j} - \rho s = \pi \ , \\
\Big( \pd{\sigma_{ji}}{x_j} + \rho f_i + \elF_i \Big) v_i + \rho \dd{u}{t} - \pd{\zeta_j}{x_j} - \rho s = \pi \ , \\
\rho \dd{u}{t} - \pd{}{x_j}\big( \zeta_j - \sigma_{ji} v_i \big) - \rho (s - f_i v_i) = \sigma_{ji}\pd{v_i}{x_j}  + \pi - \elF_i v_i\ , \\
\eqend
we have obtained the balance of internal energy. Obviously, we need to define $\pi$ and $\elF_i$ before we proceed. Among many different possibilities, the following choice leads to a thermodynamically consistent formulation. Suppose that we simply choose the electromagnetic momentum as follows
\begeq \label{electromagneto.momentum.Minkowksi}
\elmo_i = (\t \mD \times \t B)_i \ ,
\eqend
which is called \textsc{Minkowski}'s momentum. It leads to the following electromagnetic stress and force
\begeq \label{electromag.stress}
m_{ji}=-\frac12 \delta_{ji} ( H_k B_k + D_k E_k ) + H_i B_j + D_j E_i \ , \\
\elF_i = \rho z E_i + \epsilon_{ijk} J_j B_k - \epsilon_{ijk} \pd{P_j}{t} B_k - \epsilon_{ijk} P_j \pd{B_k}{t} \ ,
\eqend
after using \textsc{Maxwell}'s equations, see \ref{app.elmo.balance} for its derivation. Suppose now that we choose the electromagnetic flux as \textsc{Poynting}'s vector
\begeq
\Po_i = \epsilon_{ijk} \mH_j E_k  = \big( \t\mH \times \t E \big)_i \ ,
\eqend
in this case, as shown in \ref{app.elen.balance}, we obtain
\begeq
e^\field =   P_i E_i -B_i \MM_i + \frac12 \big( D_i E_i + H_i B_i) \comma
\pi = J_i^\fr E_i - P_i \pd{E_i}{t} + B_i\pd{\MM_i}{t} \ ,
\eqend
The production term due to the field can be rewritten by using the above definition of the electromagnetic force
\begeq
\elF_i v_i = 
\pd{}{x_j}\Big( (-P_j E_i + \MM_i B_j)v_i + \big( \t\mH \times \t E \big)_j \Big) 
+\pd{}{t} \bigg( B_i \MM_i  -P_j E_j - \frac12 D_j E_j - \frac12 B_i H_i \bigg) 
- \\
- (-P_j E_i + \MM_i B_j) \pd{v_i}{x_j}
- \EE_i \JJ_i^\fr
+ P_i \dd{E_i}{t} - B_i \dd{\MM_i}{t} 
=\pd{}{x_j}\Big( (-P_j E_i + \MM_i B_j)v_i\Big) 
+ \\
\underbrace{+\pd{\Po_j}{x_j} - \pd{e^\field}{t}}_{\displaystyle\pi} - (-P_j E_i + \MM_i B_j) \pd{v_i}{x_j}
- \EE_i \JJ_i^\fr
+ P_i \dd{E_i}{t} - B_i \dd{\MM_i}{t}.
\eqend
We refer to \cite[Sect.\,3.5]{027} for its derivation based only on subsequent use of \textsc{Maxwell}'s equations and \textsc{Maxwell--Lorentz} aether relations. Now the balance of internal energy reads
\begeq \label{bal.int.energy}
\rho \dd{u}{t} - \pd{}{x_j}\Big( \zeta_j - (\sigma_{ji} - P_jE_i + \MM_i B_j) v_i \Big) - \rho (s - f_i v_i) 
= \\
= \big(\sigma_{ji} - P_jE_i + \MM_i B_j\big)\pd{v_i}{x_j} + \EE_i \JJ_i^\fr - P_i \dd{E_i}{t} + B_i \dd{\MM_i}{t}  \ .
\eqend
We emphasize that this derivation holds for every material; we have only used one assumption and supposed that \textsc{Minkowski}'s choice in Eq.\,\eqref{electromagneto.momentum.Minkowksi} is the correct modeling for the electromagnetic momentum. Other than this assumption, the formulation is quite general such that the balance of internal energy in Eq.\,\eqref{bal.int.energy} holds for every system. Conventionally, the non-convective flux term of the internal energy is called the heat flux:
\begeq \label{energy.flux.definition}
-q_j =  \zeta_j - (\sigma_{ji} - P_jE_i + \MM_i B_j) v_i \ ,
\eqend
with the minus sign appearing because heat pumped into the system (against the surface normal) is declared as a positive work. The supply of the internal energy is a given term and is called the radiant heat:
\begeq
r = s - f_i v_i \ .
\eqend
The production term
\begeq
\Gamma = (\sigma_{ji} - P_jE_i + \MM_i B_j)\pd{v_i}{x_j} + \EE_i \JJ_i^\fr - P_i \dd{E_i}{t} + B_i \dd{\MM_i}{t} \ ,
\eqend
will be especially useful in the following section for deriving the constitutive equations. \\ 

Now by using mass balance and \textsc{Gauss}'s law, we can obtain the global forms of mass, total momentum, and internal energy balance equations in the current placement:
\begeq
\Dt{\bigg( \int_\Body \rho \d v \bigg)} = 0 \comma
\Dt{\bigg( \int_\Body \rho v_i \d v \bigg)} = \int_{\p\Body} \sigma_{ji} \d a_j + \int_\Body \rho f_i \d v + \int_\Body \elF_i \d v \ , \\
\Dt{\bigg( \int_\Body \rho u \d v \bigg)} = -\int_{\p\Body} q_i \d a_i + \int_\Body \rho r \d v + \int_\Body \Gamma \d v \ 
\eqend
These balance equations are in the current placement given in $\t x$, but we search for thermomechanical fields as functions in space $\t X$ with the reference placement, in which the mass density, displacement, and temperature are known. As the initial conditions are known, for reference placement we choose the initial placement. The volume and area elements are transformed to the initial placement by
\begeq
\d v = J \d V \comma 
\d a_j = \d A_k J (\t F^{-1})_{kj} \ , 
\eqend
with the deformation gradient and its determinant defined by 
\begeq
F_{ij} = \pd{x_i}{X_j} = \pd{u_i}{X_j} + \delta_{ij} \comma
J = \det(\t F) \ .
\eqend
Since the volume element in the initial placement is constant in time, after inserting the transformation and using \textsc{Gauss}'s law, we obtain the balance equations in a \textsc{Lagrange}an frame
\begeq \label{bal.mat.frame.local}
\rho J = \text{const.}\big|_t = \rho_0 \ , \\
\rho_0 \pd{v_i}{t} = \pd{}{X_k} \Big( J (\t F^{-1})_{kj} \sigma_{ji} \Big) + \rho_0 f_i + J \elF_i \ , \\
\rho_0 \pd{u}{t} = - \pd{}{X_k} \Big( J (\t F^{-1})_{kj} q_j \Big) + \rho_0 r + J \,\Gamma \ ,
\eqend
where each coordinate in space $\t X$ denotes a material particle and $\rho_0$ indicates the mass density of particles in the reference placement. We need constitutive equations in order to close these equations such that we can solve for the displacement and temperature.

\subsection{On the interface}

The formulation of partial differential equations generally makes the implicit assumption that all fields must be described as continuous in space and all conserved quantities are volume densities. However, this restriction is artificial, as it is perfectly valid to discuss an infinitely thin membrane with mass area density and velocity defined upon it, for example. Electromagnetism in particular has situations where this applies. Surface charges and currents are prevalent due to material discontinuities, requiring us to develop descriptions for interfaces in addition to the partial differential equations that can only be applied to ``smooth'' space. Especially between different materials, we obtain ``jump'' conditions to be satisfied in order to obtain the correct solution. \\

We name the boundary of the solid body as the interface in order to avoid a confusion to the domain boundary. The interface evolves due to the deformation. Thermomechanical fields are assumed to be computed such that the current placement of the interface is known. We will develop the equations for a generic singular surface and then restrict to the case that the singular surface is the interface.  Regions are 3D objects and surfaces are 2D objects embedded in 3D space. \\

Consider a surface $S$ and its closure $\bar S=S\cup \p S$ between two different regions $\Om^+$ and $\Om^-$ and their closures $\bar\Om^+=\Om^+\cup \p\Om^+$ and $\bar\Om^-=\Om^-\cup \p\Om^-$ that $\bar S=\bar\Om^+ \cap \bar\Om^-$. On the plus side of the surface---toward which the normal $\t n$ points---is the region $\Om^+$ and on the minus side lies $\Om^-$. We use $\Om^\pm=\Om^+\cup\Om^-$ and  denote the whole domain $\Om = \Om^\pm \cup S$ as the surface is within the domain. Its boundary is on the domain's boundary $\p S = \p\Om \cap \bar S$ such that the whole boundary reads $\p\Om=\p\Om^\pm\setminus S$. Note that the boundary of the surface, $\p S$, is a 1D loop embedded in 3D space. The surface and domain may be moving with a velocity $\t w$ and we search for a balance equation in this open system, i.e., we use a \textsc{Euler}ian frame. Now the general balance equation reads
\begeq
\Dt{\bigg( \int_{\Om^\pm} \psi^V \d v + \int_S \psi^S \d a \bigg)} 
= 
\int_{\p\Om^\pm} \Big( (w_j-v_j)\psi^V  + \Phi^V_j \Big) \d a_j + \int_{\p S} \Big( \psi^S (w_j-v_j) + \Phi^S_j \Big) \d\ell_j
+ \\
+ \int_{\Om^\pm} \big( \rho^V k^V + p^V \big) \d v + \int_{S} \big( \rho^S k^S + p^S \big) \d a \ ,
\eqend
where all volume-related quantities are denoted with a superscript $V$ and surface-related quantities with a superscript $S$. We are only interested in a special case where the surface is an interface. In other words, the surface itself is a fictitious separation without any mass area-density, $\rho^S=0$. This restriction allows the following simplification after using the geometric transformations  
\begeq
\d v = J \d V \comma
\d a_j = \d A_k J (\t F^{-1})_{kj} \comma
\d\ell_j = F_{jk} \d L_k \comma
\d a = \sqrt{\frac{g}{G}} \d A\ ,
\eqend
where $g$ and $G$ are the determinant of the surface metric tensor in the current and initial placements, respectively. After transforming to the initial placement, we obtain
\begeq
\int_{\Om^\pm} \Dt{(\psi^V J)} \d V 
= 
\int_{\p\Om^\pm} \Big( (w_j-v_j)\psi^V  + \Phi^V_j \Big) J (\t F^{-1})_{kj} \d A_k + \int_{\p S} \Phi^S_j F_{jk} \d L_k
+ \\
+ \int_{\Om^\pm} \big( \rho^V k^V + p^V \big) J \d V + \int_{S} p^S \sqrt{\frac{g}{G}} \d A \ .
\eqend
For an arbitrary field $f$, \textsc{Gauss}'s law in the initial placement reads
\begeq
\int_{\Om^\pm} \pd{f}{X_k} \d V = 
\int_{\Om^+} \pd{f}{X_k} \d V + \int_{\Om^-} \pd{f}{X_k} \d V = 
\int_{\p\Om^+} f \d A_k + \int_{\p\Om^-} f \d A_k =
\\ =
\int_{\p\Om^+\setminus S} f N_k \d A + \int_S f^+ N_k^+ \d A + \int_{\p\Om^-\setminus S} f N_k \d A + \int_S f^- N_k^- \d A =
\int_{\p\Om^\pm} f N_k \d A + \int_S \jl f N_k \jr  \d A \ ,
\eqend 
with $f^+$ or $f^-$ as the limit value from the region $\Om^+$ or $\Om^-$ on the interface and $\t N^+$ or $\t N^-$ showing outward the region $\Om^+$ or $\Om^-$, respectively. The plane normal $\t N$ is of unit length. We introduce a jump bracket notation, $\jl f N_k \jr = f^+ N_k^+ + f^- N_k^-$, by making use of the fact that a unit normal appears on both sides of the surface, $N_k=N_k^+=-N_k^-$, thus, it is possible to rewrite, $\jl f N_k \jr = (f^+ - f^- ) N_k^+$. The difference between values, $f^+ - f^-$, justifies the name of ``jump'' brackets. We assume that $\p S$ is closed (no singularities) such that we can use \textsc{Stokes}'s law with an arbitrary term $f_k$ as follows
\begeq
\int_{\p S} f_k \d L_k = \int_{S} \curl(\t f)_k \d A_k = \int_S \epsilon_{kji} \pd{f_i}{X_j} \d A_k \ .
\eqend
The general balance equation now reads
\begeq
\int_{\Om^\pm} \bigg( 
\Dt{(\psi^V J)} 
- \pd{}{X_k} \Big( \big( (w_j-v_j)\psi^V  + \Phi^V_j \big) J (\t F^{-1})_{kj} \Big) 
-\big( \rho^V k^V + p^V \big) J
\bigg) \d V 
= \\ = 
\int_S \bigg(
- \Big\jl \Big(  (w_j-v_j)\psi^V + \Phi^V_j \Big) J (\t F^{-1})_{kj} N_k \Big\jr 
+ \epsilon_{kli} \pd{\Phi^S_j F_{ji}}{X_l} N_k
+ p^S \sqrt{\frac{g}{G}}
\bigg)  \d A \ .
\eqend
The left-hand side of the latter is fulfilled within the continuum body; we need to assure that the right-hand side on the interface is satisfied as well. This restriction leads to the additional equation on the interface
\begeq
 \Big\jl \Big( (w_j-v_j)\psi^V  + \Phi^V_j \Big) J (\t F^{-1})_{kj} N_k \Big\jr 
= \epsilon_{kli} \pd{\Phi^S_j F_{ji}}{X_l} N_k
+ p^S \sqrt{\frac{g}{G}} \ .
\eqend
The volume density $\psi^V$ is a quantity per volume and its corresponding flux $\Phi^V_j$ is an area density meaning a quantity per area. Concretely, the volume density is compiled in Table~\ref{tab:supply.flux.2}: mass density, momentum density, and internal energy density. Analogously, the flux term $\Phi^S_j$ is a line density and $p^S$ is a production term; both of them exist only on the interface. For the mass, we know that neither flux nor production terms exist. Therefore, we assume that interface flux and production vanish such as
\begeq \label{sing.boundary.vel}
 \Big\jl (w_j-v_j) \rho J (\t F^{-1})_{kj} N_k \Big\jr 
=0 \ ,
\eqend
has to be fulfilled. By setting $w_i=v_i$ on the interface, we will satisfy the latter condition---in the implementation, this condition corresponds to forcing the computational mesh to follow the interface. There is, however, a flux term for the momentum. In mechanics the surface tension is the stress on the interface, see \cite{cammarata1994}, \cite{vermaak1968}, \cite{mays1968}, \cite{shuttleworth1950}. In the case of electromagnetism, the production term on the interface needs to be calculated. As the volumetric production term is given in Eq.\,\eqref{electromag.stress}$_2$, the area production term is due to the surface charges and currents. Surface charges and currents become important in mixtures with adhesion between particles. For a solid body, we simply neglect these effects and obtain the balance of momentum,
\begeq
 \Big\jl \Big(  (w_j-v_j) \rho v_i + \sigma_{ji} \Big) J (\t F^{-1})_{kj} N_k \Big\jr 
=0 \ .
\eqend
Since $\t w = \t v$ on the interface, the first term drops out leaving 
\begeq \label{sing.tm1}
\Big\jl \sigma_{ji} J (\t F^{-1})_{kj} N_k \Big\jr 
=0 \ .
\eqend
Analogously, for the balance of internal energy on the interface, we use $\t v = \t w$ and neglect the production as well as the line density flux term (surface heat) such that we obtain
\begeq \label{sing.tm2}
\big\jl q_j N_k J (\t F^{-1})_{kj} \big\jr 
=0 \ .
\eqend
We may discuss the displacement field as a continuous variable, meaning that the continuum body has no cracks. An analogous argument yields that all primitive variables are continuous and will thus be modeled as such. Moreover, the singular surface is an interface in which the surface normal in the current placement has to be continuous. These assertions of continuity yield the following conditions:
\begeq
A_i^+ = A_i^-  \comma
\phi^+ = \phi^- \comma
u_i^+ = u_i^- \comma
T^+ = T^- \comma
\jl n_j \jr = \jl N_k J (\t F^{-1})_{kj}  \jr = 0  \ .
\eqend
In the numerical implementation we need to guarantee that the latter equations apply on the interface. This condition is equivalent imposing that all primitive variables---$\t A$, $\phi$, $\t u$, and $T$---are (piecewise) continuous within the whole computational domain. Moreover, by deriving \textsc{Maxwell}'s equations from the balance equations, the following equations are obtained
\begeq \label{sing.em1}
\jl n_i \mD_i \jr = 0 \comma \jl \epsilon_{ijk} n_j \mH_k \jr = 0 \ ,
\eqend
under the assumption that no surface charges and currents exist. The jump terms emerge due to the difference of material parameters between the adjacent materials. As we use continuous $\t A$, $\phi$ in combination with Eqs.\,\eqref{em.DHrelations}, \eqref{em.MLrelations}, we can also deduce 
\begeq \label{sing.em2}
\jl n_i P_i \jr = 0 \comma \jl \epsilon_{ijk} n_j \MM_k \jr = 0 \ .
\eqend 

\section{Constitutive equations}
\label{sec:constitutive}
Various thermodynamical procedures exist in the literature. They all aim at deriving the constitutive equations for $\tt\JJ^\fr$, $\t\sigma$, $\t q$, $\t P$, and $\tt\MM$. We use a similar strategy as in \cite{groot1984} in which the main assumption is that the internal energy is recoverable, leading to an entropy with primary variables but not fluxes. This assumption is a limitation in the theory; for extension see \cite{jou1999}, \cite{muller2013}. Moreover, we neglect irreversible effects in polarization such as hysteresis. The presented theory incorporates only elastic materials; in other words, we neglect irreversible deformation called plasticity. Only first gradient of the primitive variables are considered; for higher gradient theories, we refer to \cite{altenbach2013generalized}, \cite{abali2017highergrad}. \\

We will compute the primitive variables in the whole domain: the temperature $T$, displacement $\t u$, electric potential $\phi$, and magnetic potential $\t A$. In the end, every proposed constitutive equation has to depend only on the primitive variables, including their space and time derivatives. Since we have general relations in Eq.\,\eqref{em.ansatz} between electromagnetic fields, $\t E$, $\t B$, and electromagnetic potentials, $\phi$, $\t A$, we can also define dependencies with respect to $\t E$ and $\t B$. The constitutive equations are necessary where a material occupies a region, they are also called material equations and are defined in the reference (herein initial) placement. We start with Eq.\,\eqref{bal.mat.frame.local}$_3$, i.e., the balance of internal energy
\begeq \label{int.energy.mat.frame}
\rho_0 \pd{u}{t} = - \pd{Q_i}{X_i} + \rho_0 r + J\Gamma \comma
Q_i = J (\t F^{-1})_{ij} q_j \ , \\
\Gamma = \Xi_{ji} \pd{v_i}{x_j} + \EE_i \JJ_i^\fr - P_i\dd{E_i}{t} + B_i \dd{\MM_i}{t} \comma 
\Xi_{ji} = \sigma_{ji} - P_j E_i + \MM_i B_j \ .
\eqend
In the following we will define constitutive equations for $u$, $\t Q$, $\t\sigma$, $\tt\JJ^\fr$, $\t P$, $\tt\MM$. The supply term, $r$, is a given function in space and time. Production of internal energy, $\Gamma$, consists of four terms. The third and fourth terms are crucial for deriving the constitutive equations for $\t P$ and $\tt\MM$. The second term in the production of internal energy is called \textsc{Joule}'s heat. We will see this term as a purely dissipative phenomenon. The first term can be rewritten
\begeq
J \Xi_{ji} \pd{v_i}{x_j} = J \Xi_{ji} \pd{v_i}{X_k} (\t F^{-1})_{kj} = N_{ki} \pd{v_i}{X_k} \ , 
\eqend
with the nominal stress $N_{ji} = J (\t F^{-1})_{jk} \Xi_{ki}$ defined in the initial placement. Moreover, we observe
\begeq
\dd{ F_{ij}}{t} = \dd{}{t}\pd{x_i}{X_j} = \pd{^2 x_i}{t \p X_j} = \pd{^2 x_i}{X_j \p t} = \pd{v_i}{X_j} \ ,
\eqend
such that the balance of internal energy becomes
\begeq
\rho_0 \pd{u}{t} = - \pd{Q_i}{X_i} + \rho_0 r + N_{ji} \dd{F_{ij}}{t} + J \EE_i \JJ_i^\fr - J P_i\dd{E_i}{t} + J B_i \dd{\MM_i}{t} \ .
\eqend
Now we make several assumptions in order to deduce an equation for equilibrium. Since fields do not change in equilibrium, any external supply such as $r$ or production such as \textsc{Joule}'s heat vanishes in equilibrium. In the most general case, we have to assume that the internal energy, heat flux, and electromagnetic polarization have reversible and irreversible parts. First, we assume that the internal energy is fully recoverable, neglecting the irreversible part of the internal energy. This is a conventional method that is appropriate for many engineering systems. We ignore effects of flux rates (the stress rate and heat flux rate) into the internal energy. For systems with high temperature rates or high velocity gradients, the proposed method would be inaccurate and extended thermodynamics deals with this issue, we refer to \cite{jou1999}, \cite{muller2013}. The reversible part of the heat flux is given by a term called specific entropy, $\eta$, see (\cite[Ch.\,XIV, \textsection2]{groot1984})---this definition goes back to \cite{Caratheodory1909}---as follows
\begeq
- \pd{Q_i}{X_i} = \rho_0 T \dd{\eta}{t}  \ ,
\eqend
where we need a constitutive equation for the entropy. We neglect any hysteresis in the electromagnetic response and assume that electric and magnetic polarizations are reversible. In equilibrium, the balance of internal energy reads
\begeq
\rho_0 \d u = \rho_0 T \d \eta + N_{ji} \d F_{ij} - J P_i\d E_i + J B_i \d \MM_i \ , \\
\d u = T \d \eta + \frac1{\rho_0} N_{ji} \d F_{ij} - \frac1{\rho} P_i\d E_i + \frac1{\rho} B_i \d \MM_i \ , \\
\d u = T \d \eta + n_{ji} \d F_{ij} - p_i\d E_i +  B_i \d \mm_i - B_i \MM_i \d\v \ , \\
\eqend 
where  $p_i=P_i/\rho$ is the specific electric polarization in the current placement,  $\mm_i=\MM_i/\rho$ is the specific magnetic polarization in the current placement, $\pre = B_i\MM_i$ is the electromagnetic pressure (owing to its unit), $\v=1/\rho$ is the specific volume (volume per mass), and  $n_{ji}=N_{ji}/\rho_0$ is the specific nominal stress in the initial placement. The last line can be called \textsc{Gibbs}'s equation. It is a \textit{perfect differential} that allows us to determine the internal energy by integrating its differential form. This assumption is called the first law of thermodynamics, see \cite{pauli1973}. From the differential form, we immediately realize that the internal energy depends on $\{\eta, F_{ij}, E_i, \mm_i, \v\}$. Since we want to define $\eta$ and $\mm_i$, it is beneficial to have a dependence on their conjugate variables, namely $T$ and $B_i$. This is achieved by introducing a free energy:
\begeq
\free = u - T\eta - B_i \mm_i \ ,
\eqend
and assuming that a perfect differential of this free energy exists such that 
\begeq
\d\free = \d u  -\d T \eta - T \d\eta - \d B_i \mm_i - B_i \d\mm_i  \ , \\
\d\free = -\eta \d T + n_{ji} \d F_{ij} - p_i \d E_i - \mm_i \d B_i - \pre \d\v \ .
\eqend
The free energy depends on $\{ T, F_{ij}, E_i, B_i, \v \}$. They can be called \textit{primary} or \textit{state} variables. We have the following obvious relations 
\begeq
\pd{\free}{T} = -\eta \comma
\pd{\free}{F_{ij}} = n_{ji} \comma
\pd{\free}{E_i} = -p_i \comma
\pd{\free}{B_i} = -\mm_i \comma
\pd{\free}{\v} = -\pre = -B_i\MM_i \ .
\eqend
Additionally, because $\free$ depends on primary variables, each so-called \textit{dual} variable, $\{ \eta, n_{ji}, p_i, \mm_i, \pre \}$, depends on the same set of arguments---this is often named as equipresence principle, see \cite[\textsection 293.$\eta$]{truesdell_toupin}---leading to
\begeq
\d\eta = \tilde c^{11} \d T + \tilde c^{12}_{kl} \d F_{kl} + \tilde c^{13}_k \d E_k + \tilde c^{14}_k \d B_k + \tilde c^{15} \d \v
\ , \\
\d n_{ji} = \tilde c^{21}_{ji} \d T + \tilde c^{22}_{jikl} \d F_{kl} + \tilde c^{23}_{jik} \d E_k + \tilde c^{24}_{jik} \d B_k + \tilde c^{25}_{ji} \d \v
\ , \\
\d p_i = \tilde c^{31}_{i} \d T + \tilde c^{32}_{ikl} \d F_{kl} + \tilde c^{33}_{ik} \d E_k + \tilde c^{34}_{ik} \d B_k + \tilde c^{35}_i \d \v
\ , \\
\d\mm_i =\tilde c^{41}_{i} \d T + \tilde c^{42}_{ikl} \d F_{kl} + \tilde c^{43}_{ik} \d E_k + \tilde c^{44}_{ik} \d B_k + \tilde c^{45}_i \d \v
\ , \\
\d \pre =\tilde c^{51} \d T + \tilde c^{52}_{kl} \d F_{kl} + \tilde c^{53}_k \d E_k + \tilde c^{54}_k \d B_k + \tilde c^{55} \d \v
\ . 
\eqend
All material parameters, $\tilde c^{11}$, $\tilde c^{12}$, $\dots$, and $\tilde c^{55}$, need to be measured independently. There is a reduction of measurements due to the \textsc{Maxwell} symmetry relations (see \ref{maxwell.symmetry} for their derivations). We can rewrite the above constitutive equations in the linear algebra fashion by using block matrices for the sake of clarity by 
\begeq \label{dual.var1}
\begin{pmatrix} \d\eta \\ \d n_{ji} \\ \d p_i \\ \d\mm_i \\ \d\pre \end{pmatrix}
=
\begin{pmatrix} 
\tilde c^{11} & -\tilde c^{21}_{lk} & \tilde c^{31}_{k} & \tilde c^{41}_{k} & \tilde c^{51} \\
\tilde c^{21}_{ji} & \tilde c^{22}_{jikl} & -\tilde c^{32}_{kij} & -\tilde c^{42}_{kij} & -\tilde c^{52}_{ij} \\
\tilde c^{31}_i & \tilde c^{32}_{ikl} & \tilde c^{33}_{ik} & \tilde c^{43}_{ki} & \tilde c^{53}_i \\
\tilde c^{41}_i & \tilde c^{42}_{ikl} & \tilde c^{43}_{ik} & \tilde c^{44}_{ik} & \tilde c^{54}_i \\
\tilde c^{51} & \tilde c^{52}_{kl} & \tilde c^{53}_{k} & \tilde c^{54}_{k} & \tilde c^{55} 
\end{pmatrix}
\begin{pmatrix} \d T \\ \d F_{kl} \\ \d E_k \\ \d B_k \\ \d\v \end{pmatrix} .
\eqend\\

The experiments to determine these coefficients are established by varying a primary variable while holding all other primary variables constant and measuring one dual variable. Consider the first dual variable: instead of measuring entropy, we observe the heat flux $\del Q = T\d\eta$ and then measure temperature by holding the deformation gradient, electromagnetic fields, and mass density fixed at a specific value such that their variations are zero, i.e., $\d F_{kl}=0$, $\d E_i=0$, $\d B_i=0$, $\d\v=0$. The parameter $c$ relating heat flux to temperature is called the specific heat capacity, $\del Q=c\d T$, so we obtain $\tilde c^{11}=c/T$. Heat capacity can depend on fields besides the temperature. The specific stiffness\footnote{Note that this stiffness tensor is not the stiffness tensor which is normally discussed. In our derivation, we have a stiffness tensor to be the derivative of a non-symmetric tensor with respect to another non-symmetric tensor. Thus, the minor symmetries are not present. Further, since we never declared a quadratic strain energy function, the major symmetry is also not necessarily present. The symmetry relations are introduced as a consequence of any existing crystal symmetries. For example, in the case of an isotropic material, all aforementioned symmetries arise.} tensor, $\tilde c^{22}_{jikl}$, is measured for a specifically chosen state of a held temperature, electromagnetic fields, and mass density and then applying a deformation and measuring stress. All other coefficients can be measured in analogous settings. For example $\tilde c^{33}_{ik}$ and $\tilde c^{44}_{ik}$ are susceptibilities. Between the electromagnetic pressure $\pre$ and specific volume $\v$, the coefficient $\tilde c^{55}$ can be measured by applying magnetic field and measuring the volume change. Such measurements are very challenging, but possible. Moreover, the off-diagonal terms are different coupling terms between primary variables. For example, $\tilde c^{23}_{jik}$ and $\tilde c^{24}_{jik}$ are coupling stress with electromagnetic fields, called the specific piezoelectric and piezomagnetic material coefficients, respectively. \\ 

Finding the suggested coefficients in the literature is challenging due to difficulties in making the described measurements. Hence, we will switch to quantities that are more regularly measured and thus appear more frequently in the literature. Coefficients of thermal expansion, $\alpha_{ij}$, are measured by varying temperature and measuring length change 
\begeq
\d F_{ij} = \alpha_{ij} \d T \ ,
\eqend
by holding every other variable fixed such that we acquire from Eq.\,\eqref{dual.var1}$_{2,3,4,5}$ the following relations 
\begeq
0 = \tilde c^{21}_{ji} \d T + \tilde c^{22}_{jikl} \alpha_{kl} \d T \Rightarrow \tilde c^{21}_{ji} = -\tilde c^{22}_{jikl} \alpha_{kl} \ , \\
0 = \tilde c^{31}_{i} \d T + \tilde c^{32}_{ikl} \alpha_{kl} \d T \Rightarrow \tilde c^{31}_{i} = -\tilde c^{32}_{ikl} \alpha_{kl} \ , \\
0 = \tilde c^{41}_{i} \d T + \tilde c^{42}_{ikl} \alpha_{kl} \d T \Rightarrow \tilde c^{41}_{i} = -\tilde c^{42}_{ikl} \alpha_{kl} \ , \\
0 = \tilde c^{51} \d T + \tilde c^{52}_{kl} \alpha_{kl} \d T \Rightarrow \tilde c^{51} = -\tilde c^{52}_{kl} \alpha_{kl} \ .
\eqend
The demonstrated procedure is well-known in thermodynamics, see for example \cite{nye1969}. We emphasize that all material coefficients need to be determined experimentally and they may depend on all state variables, $\{T, F_{ij}, E_i, B_i, \v\}$. Of course such experiments are very challenging and often the coefficients are determined by using an inverse analysis. Some nonlinear phenomena are captured in experiments, for example heat capacity or thermal expansion coefficients depending on temperature, stiffness tensor depending on deformation gradient (called hyperelasticity), electric or magnetic susceptibility depending on electric field or magnetic flux. We will concretely demonstrate the case with linear equations and nonlinear equations in the following. The real measurable is the energy such that a formulation based on the energy is very useful---this configuration is the case in nonlinear equations.

\subsection{Linear equations at equilibrium}
 
Firstly, we present the case in which material coefficients are constants in the corresponding variable of integration: $c$ is constant in $T$ and $\tilde c^{22}_{jikl}$ is constant in $F_{ij}$ and so on. In this case, we can easily integrate from a reference state $T=T_\Ref$, $F_{ij}=\delta_{ij}$, $E_i=0$, $B_i=0$, and $\v=\v_\Ref$  to the current state. We define the reference state as the state in which all dual variables vanish. After integrating and multiplying by mass density, we obtain the relations 
\begeq
\eta = c \ln\Big(\frac{T}{T_\Ref}\Big) + \tilde c^{22}_{lkij} \alpha_{ij} (F_{kl}-\delta_{kl}) - \tilde c^{32}_{kij} \alpha_{ij} E_k - \tilde c^{42}_{kij} \alpha_{ij} B_k - \tilde c^{52}_{ij} \alpha_{ij} (\v-\v_\Ref) \ , \\
N_{ji} = -\rho_0 \tilde c^{22}_{jikl} \alpha_{kl} (T-T_\Ref) + \rho_0 \tilde c^{22}_{jikl} (F_{kl}-\delta_{kl}) - \rho_0 \tilde c^{32}_{kij} E_k - \rho_0 \tilde c^{42}_{kij} B_k - \rho_0 \tilde c^{52}_{ij} (\v-\v_\Ref) \ , \\ 
P_i = -\rho \tilde c^{32}_{ikl} \alpha_{kl} (T-T_\Ref) + \rho \tilde c^{32}_{ikl} (F_{kl}-\delta_{kl}) + \rho \tilde c^{33}_{ik} E_k + \rho \tilde c^{43}_{ki} B_k + \rho \tilde c^{53}_{i} (\v-\v_\Ref) \ , \\  
\MM_{i} = -\rho \tilde c^{42}_{ikl} \alpha_{kl} (T-T_\Ref) + \rho \tilde c^{42}_{ikl} (F_{kl}-\delta_{kl}) + \rho \tilde c^{43}_{ik} E_k + \rho \tilde c^{44}_{ik} B_k + \rho \tilde c^{54}_{i} (\v-\v_\Ref) \ , \\ 
\pre = -\tilde c^{52}_{kl} \alpha_{kl} (T-T_\Ref) + \tilde c^{52}_{kl} (F_{kl}-\delta_{kl}) + \tilde c^{53}_{k} E_k + \tilde c^{54}_{k} B_k + \tilde c^{55} (\v-\v_\Ref) \ .
\eqend 
The reference temperature $T_\Ref$ as well as the specific volume $\v_\Ref$ need to be specified. Although these values are material specific, we assume that the initial mass density and temperature can be chosen as reference values. In other words, we start simulating from the ground state without entropy, stress, and polarization. Since the deformation gradient is the sum of the identity and the displacement gradient, we have $\p u_i/\p X_j = F_{ij}-\delta_{ij}$. It is important to mention that we use $\p u_i/\p X_j$ instead of strain because the additional term in the nominal stress fail to be symmetric. The mechanical part of the nominal stress (transpose of the \textsc{Piola} stress) relates the current force to initial area and is often called the \textit{engineering} stress. In an experiment, the length change is recorded and often the so-called \textit{engineering} strain is reported as current length (measured length change plus initial length) divided by the initial length, which is one component of $F_{ij}$. Hence, we can use the reported stiffness values in $\tilde c^{22}_{ijkl}$. For the thermodynamic analysis we have used specific variables; however, in practice, the measurements are established utilizing stress and polarization, $N_{ji}$, $P_i$, $\MM_i$, instead of their per mass values, $n_{ji}$, $p_i$, $\mm_i$. Therefore, we rename the material coefficients as follows
\begeq
\rho_0 \tilde c^{22}_{jikl} = C_{jikl} \comma
\rho_0 \tilde c^{32}_{kij} = \tilde T_{kij} \comma
\rho_0 \tilde c^{42}_{kij} = \tilde S_{kij} \ , \\
\rho \tilde c^{33}_{ik} = \eps_0 \chi_{ik}^\el \comma
\rho \tilde c^{43}_{ki} = \tilde R_{ki} \comma
\rho \tilde c^{44}_{ik} = (\t\mu_\ma^{-1})_{ij} \chi_{jk}^\ma  \ .
\eqend
As we have used the electromagnetic pressure (energy density) $\pre=\MM_i B_i$ we obtain by multiplying $\MM_i$ by $B_i$ and matching the coefficients to $\pre$, 
\begeq
\tilde c^{52}_{kl} = \rho \tilde c^{42}_{ikl} B_i = J^{-1} \tilde S_{ikl} B_i \comma
\tilde c^{53}_{k} = \rho \tilde c^{43}_{ik} B_i = \tilde R_{ik} B_i \ , \\
\tilde c^{54}_{k} = \rho \tilde c^{44}_{ik} B_i = (\t\mu^{-1}_\ma)_{ij} \chi_{jk}^\ma B_i \comma
\tilde c^{55}_{k} = \rho \tilde c^{54}_{i} B_i  \ .
\eqend
Finally, we acquire the following constitutive equations
\begeq \label{linear.eqs}
\eta = c \ln\Big(\frac{T}{T_\Ref}\Big) + \v_0 C_{lkij} \alpha_{ij} \pd{u_k}{X_l} - \v_0 \tilde T_{kij} \alpha_{ij} E_k - \v_0 (2-J^{-1}) \tilde S_{kij} \alpha_{ij} B_k \ , \\
N_{ji} = C_{jikl} \Big( -\alpha_{kl} (T-T_\Ref) +  \pd{u_k}{X_l} \Big) - \tilde T_{kij} E_k - (2-J^{-1}) \tilde S_{kij} B_k  \ , \\ 
P_i = J^{-1} \tilde T_{ikl} \Big( -\alpha_{kl} (T-T_\Ref) + \pd{u_k}{X_l} \Big) + \eps_0 \chi^\el_{ik} E_k + (2-J^{-1}) \tilde R_{ki} B_k  \ , \\  
\MM_{i} = J^{-1} \tilde S_{ikl} \Big( -\alpha_{kl} (T-T_\Ref) + \pd{u_k}{X_l} \Big) + \tilde R_{ik} E_k + (2-J^{-1}) (\t\mu^{-1}_\ma)_{ij} \chi^\ma_{jk} B_k \ .
\eqend 
Especially, the stiffness tensor $C_{ijkl}$, electric susceptibility $\chi^\el_{ij}$, magnetic susceptibility $\chi^\ma_{ij}$, permeability of the vacuum $\eps_0$, and permittivity of the material $\mu_{ij}$ are available for many engineering materials. Piezoelectric and piezomagnetic coefficients, $\tilde T_{ijk}$ and $\tilde S_{ijk}$, are also possible to find. Often, as stated in \cite{meitzler1988}, the measurements are undertaken by varying electric field and measuring displacement gradient, $\d \p u_i/\p X_j = d_{kij} \d E_k$, and by determining $d_{kij}$ with the standard notation. In this case, we can readily find from Eq.\,\eqref{dual.var1}$_2$ the relation 
\begeq
\tilde T_{mij} = C_{ijkl} d_{mkl} \ .
\eqend
Also quite often, the piezomagnetic constants are given in T$\hat =$N/(A\,m) as $\tilde q_{ijk} = \mu_0 \tilde S_{ijk}$. The magnetoelectric coupling $\tilde R_{ij}$ is rarely measured. \\

\subsection{Nonlinear equations at equilibrium}

Secondly, we present the case in which the constitutive equations at equilibrium are modeled by using nonlinear relations. In this case, by starting with Eq.\,\eqref{dual.var1}, we can integrate and define the constitutive equation if the functional form of the coefficients are explicitly known. In a slightly abusive notation, we can include the nonlinear materials, for example for hyperelasticity by using a $C_{jikl}$ that depends on $F_{ij}$ and by recalling
\begeq \label{nonlin.eqs}
\pd{\free}{F_{ij}} = n_{ji} \comma 
\pd{n_{ji}}{F_{kl}} = \tilde c^{22}_{jikl} \comma
 \rho_0 \tilde c^{22}_{jikl} = C_{jikl}  \comma
\pd{n_{ji}}{B_{k}} = \tilde c^{24}_{jik} = -\tilde c^{42}_{kij} \comma
 \rho_0 \tilde c^{42}_{kij} = \tilde S_{kij} \ ,
\eqend
to obtain
\begeq
C_{jikl} = \rho_0 \pd{^2 \free}{F_{ij} \p F_{kl}} \comma
\tilde S_{kij} = -\rho_0 \pd{^2 \free}{B_k \p F_{ij}} \ .
\eqend
Especially for soft materials, the measurement of the free energy is more feasible than the stiffness tensor; we refer to experiments in \cite{treloar1975}. For the case of an isotropic material, the free energy's dependency on the deformation gradient can be stated in terms of the invariants. From the thermodynamics point of view, any function can be suggested as a material equation. There are some restrictions because of the approximative computation in a weak form. These restrictions are called \textit{ellipticity}---in the special case of an isotropic material, \textit{invertibility}---in order to assure a smooth deformation gradient, $\t F^+ =\t F^-$, within the computational domain as proven in \cite{rosakis1990}. For the isotropic case, the dependency is given by invariants instead of a single argument $\t F$. Then, ellipticity holds in every argument of the free energy. This case is called \textit{quasiconvexity}. When designing a constitutive response, the fundamental form and material constants are chosen to hold quasiconvexity in order to compute the primitive variables with sufficient smoothness.

\subsection{Equations at non-equilibrium}
Since we have now defined all necessary constitutive equations, we insert \textsc{Gibbs}'s equation in the balance of internal energy to acquire
\begeq
\rho_0 T \pd{\eta}{t} = - \pd{Q_j}{X_j} + \rho_0 r + J \EE_i \JJ^\fr_i \ . \\
\eqend
We emphasize that we have assumed elasticity and no dissipation in polarization. With a slight rearrangement, we obtain the balance of entropy:
\begeq \label{bal.entropy}
\rho_0 \pd{\eta}{t} + \pd{}{X_j}\Big( \frac{Q_j}{T} \Big) -\rho_0 \frac{r}{T} = - \frac{Q_j}{T^2} \pd{T}{X_j} + \frac{J}{T} \EE_i \JJ_i^\fr \ 
\eqend
The entropy flux can differ from this formulation if the energy flux in Eq.\,\eqref{energy.flux.definition} is defined differently. A well-known alternative includes the term $\t E \times \t\mH$ into the heat flux. In this formulation, the entropy flux and production would have an additional term in the heat flux similar to a \textsc{Hall} effect; for its elaborate discussion see \cite[\textsection9.9.4]{mueller1985}. We continue by using the chosen definition leading to the usual definition as above and declare the thermodynamical fluxes $-Q_i$ and $\JJ_i^\fr$ to depend on the thermodynamical forces $T^{-2} \p T/\p X_i$ and $J T^{-1} \EE_i$. By using representation theorems, we can determine the following functional relationships
\begeq
-Q_i = \tilde k^{11} T^{-2} \pd{T}{X_i} + \tilde k^{12} J T^{-1} \EE_i \comma
\JJ^\fr_i = \tilde k^{21} T^{-2} \pd{T}{X_i} + \tilde k^{22} J T^{-1} \EE_i \ .
\eqend  
These relations are the most general constitutive equations with coefficients $\tilde k^{\times\times}$ as scalar functions depending on (the invariants of) the temperature gradient and electric field. According to the second law of thermodynamics, the entropy production has to be positive for any process, i.e.,
\begeq
\tilde k^{11} T^{-4} \pd{T}{X_i} \pd{T}{X_i} + (\tilde k^{12} + \tilde k^{21}) J T^{-3} \EE_i \pd{T}{X_i} + \tilde k^{22} J T^{-1} \EE_i \EE_i \geq 0 
\eqend
such that we obtain the restrictions
\begeq \label{constitutive.coeffs.noneq}
\tilde k^{11} \geq 0 \comma
\tilde k^{12} + \tilde k^{21} = 0 \comma
\tilde k^{22} \geq 0 \ ,
\eqend
since the absolute temperature is positive, $T>0$, and the determinant of the deformation gradient is positive, $J>0$. The coefficients, $\tilde k^{\times\times}$, are scalar functions. For the simplified case of constant coefficients, the above linear relations can be derived by using statistical mechanics, where the second restriction $\tilde k^{12} = - \tilde k^{21}$ is known as \textsc{Onsager} relations. For showing the relevance to well-established phenomenological equations, we rename the material parameters
\begeq \label{constitutive.coeffs.rename}
\kappa = \tilde k^{11} T^{-2} \comma 
\varsigma = \tilde k^{22} J T^{-1} \comma 
\varsigma \pi = -\tilde k^{12} T^{-2} \ ,
\eqend
and obtain the following constitutive equations:
\begeq \label{constitutive.heatflux.current}
Q_i = -\kappa \pd{T}{X_i} + \varsigma \pi T J \EE_i \comma 
\JJ_i^\fr = \varsigma \pi \pd{T}{X_i} + \varsigma  \EE_i \ .
\eqend
The heat conduction coefficient, $\kappa$, electrical conductivity, $\varsigma$, and the thermoelectric coupling coefficient, $\pi$, are determined experimentally. The thermoelectric coupling is found to be constant for many engineering materials. Often it is called the \textsc{Peltier} constant. Although every conducting material possesses a \textsc{Peltier} constant, it might be small enough to be ignored. For the case of $\kappa=\text{const.}$ and $\pi=0$, the constitutive equation for the heat flux is called \textsc{Fourier}'s law. For the case of $\varsigma=\text{const.}$ and $\pi=0$, the constitutive equation for the electric current is named after \textsc{Ohm}. We stress that the second law is fulfilled as a consequence of Eqs.\,\eqref{constitutive.coeffs.noneq}, \eqref{constitutive.coeffs.rename} by having $\kappa\geq 0$ and $\varsigma\geq 0$. For the thermoelectric constant $\pi$ there is no such restriction as we have used it in both fluxes with different signs. There are no assumptions or simplifications in this theoretical derivation of $\t Q$ and $\tt\JJ^\fr$. In applications we will use linear constitutive equations by setting $\kappa$, $\varsigma$, $\pi$ constants as we fail to find their experimental determination depending on invariants of temperature gradient and electric field in the literature.
\section{Computational approach}
\label{sec:computational}
A considerable amount of studies and efforts are undertaken for solving mechanics and electromagnetism. For thermomechanics, we may claim the finite element method (FEM) with \textsc{Galerkin} approach using standard continuous piecewise polynomials called $\mathscr P_n$ elements, which are $n$-times differentiable, is \textit{the gold standard}. If it comes to electromagnetism, there are several ``different'' methods among scientists---for example see \cite{bossavit1988}, \cite{jiang1998least}, \cite{ciarlet1999}, \cite{sadiku2000numerical}, \cite{hiptmair2002finite}, \cite{bastos2003electromagnetic}, \cite{monk2003finite}, \cite[Sect.\,17]{demkowicz2006}, \cite{gibson2007method}, \cite{li2009}, \cite{gillette2016}, \cite[Sect. 3]{027}---and a consensus as to the ``best'' approach is yet missing. If one aims at solving electromagnetic fields $\t E$ and $\t B$ by satisfying \textsc{Maxwell}'s equations, then FEM with standard elements cannot be used and there are various so-called mixed elements, see \cite{arnold2014}, whose techniques are based on works of \cite{raviart1977} and \cite{nedelec1980}. Very roughly summarized, the mixed elements possess special forms for the functions fulfilling two of \textsc{Maxwell}'s equations. From a theoretical point of view, this method is correct since the ultimate goal is to compute the electromagnetic fields directly. \\

As we have seen in the formulation, the introduction of electromagnetic potentials, $\phi$ and $\t A$, simplifies the procedure by solving two of \textsc{Maxwell}'s equations in a closed form. As a consequence, we can set the objective to compute the electromagnetic potentials by means of standard elements. This procedure is implemented in \cite[Chap.\,3]{027} and its accuracy of the computation is demonstrated in \cite{abali2018}. After computing electromagnetic potentials, we can easily derive the electromagnetic fields by post-processing the solution with the equations 
\begeq \label{E.B.fields.in.X}
E_i = -\pd{\phi}{x_i} - \pd{A_i}{t} \comma
B_i = \epsilon_{ijk} \pd{A_k}{x_j} \ ,
\eqend
where $\t x$ denotes the coordinate in the \textsc{Euler}ian frame; i.e., the derivatives $\partial/\partial t$ and $\partial / \partial x_i$ are taken with respect to the spatial position in the laboratory frame. The space in this frame will be discretized for the computation by using standard triangulation/tetrahedralization methods. The nodes possess coordinates in the laboratory frame. However, the nodes can move with an arbitrary velocity $\t w$ which we call the domain or mesh velocity. The free choice of the arbitrary $\t w$ will be exploited for mesh quality. In order to satisfy Eq.\,\eqref{sing.tm1}, we choose $\t w$ based on the motion of the continuum body. All script values, $\tt\DD$, $\tt\HH$, $\tt\EE$, and $\tt\BB$ are measured in the moving domain. We will use their counterparts $\t D$, $\t H$, $\t E$, and $\t B$ in the laboratory frame (as a coincidence $\tt\DD=\t D$ and $\tt\BB=\t B$). The electromagnetic potentials $\phi$ and $\t A$ are calculated by fulfilling the governing Eqs.\,\eqref{governing.em} with the total charge:
\begeq
J_i = \JJ_i^\fr + q^\fr v_i + \pd{P_i}{t} + \epsilon_{ijk} \pd{\MM_k}{x_j} \comma
q^\fr = \pd{\mD_i}{x_i} \ , \\
D_i = \eps_0 E_i \comma H_i = \frac1{\mu_0} B_i \comma
\mD_i = D_i + P_i \comma \mH_i = H_i - \MM_i \ , \\
\eqend
as well as the material specific relations:
\begeq \label{current.polarization}
\JJ_i^\fr = \varsigma \pi \pd{T}{X_i} + \varsigma  \EE_i \ , \\
P_i = J^{-1} \tilde T_{ikl} \Big( -\alpha_{kl} (T-T_\Ref) + \pd{u_k}{X_l} \Big) + \eps_0 \chi^\el_{ik} E_k + (2-J^{-1}) \tilde R_{ki} B_k  \ , \\  
\MM_{i} = J^{-1} \tilde S_{ikl} \Big( -\alpha_{kl} (T-T_\Ref) + \pd{u_k}{X_l} \Big) + \tilde R_{ik} E_k + (2-J^{-1}) (\t\mu^{-1}_\ma)_{ij} \chi^\ma_{jk} B_k \ ,
\eqend
where the differentiation in space occurs in the \textsc{Lagrange}an frame in coordinates $\t X$, which can be visualized as moving with the material velocity $v_i = \p u_i / \p t$. The displacement, $\t u$, and temperature, $T$, as well as their gradients are computed in the \textsc{Lagrange}an frame. For example, $\p u_i/\p X_j$ is computed in the \textsc{Lagrange}an frame and then mapped to the \textsc{Euler}ian frame by projecting this tensor to each coordinate $\t x$. The deformation is obtained by fulfilling the balance of momentum
\begeq \label{governing.u}
\rho_0 \pd{^2 u_i}{t^2} = \pd{}{X_k} \Big( J (\t F^{-1})_{kj} \sigma_{ji} \Big) + \rho_0 f_i + J \elF_i
\eqend
by phrasing the displacement $\t u$ as a function in the initial placement. Initial mass density, $\rho_0$, is a given constant for a homogeneous material and is a known function in $\t X$ for a heterogeneous material, as is consequently the specific body force due to the gravitation, $\t f$. The electromagnetic force density in the \textsc{Lagrange}an frame reads
\begeq
J \elF_i = J q E_i + J \epsilon_{ijk} J_j B_k - J \epsilon_{ijk} \pd{P_j}{t} B_k - J \epsilon_{ijk} P_j \pd{B_k}{t} \ ,
\eqend
where this time $q=\p D_i/\p x_i$, $B_i$ and $E_i$ from Eq.\,\eqref{E.B.fields.in.X} are computed in the \textsc{Euler}ian frame and projected to the \textsc{Lagrange}an frame as scalar, vectors, respectively. The stress is given by
\begeq
\sigma_{ji} = J^{-1} F_{jk} N_{ki} + P_j E_i - \MM_i B_j \ , 
\eqend
with the nominal stress, $N_{ij}$, obtained in Eq.\,\eqref{linear.eqs}$_2$ for a linear or in Eq.\,\eqref{nonlin.eqs}$_1$ for a nonlinear material. Hence, Eq.\,\eqref{governing.u} is closed and can be solved. Temperature, $T$, can either be solved by using the balance of internal energy or the balance of entropy. We use the latter 
\begeq \label{governing.entropy}
\rho_0 \pd{\eta}{t} + \pd{}{X_j} \Big( \frac{Q_j}{T} \Big) - \rho_0 \frac{r}{T} = -\frac{Q_j}{T^2} \pd{T}{X_j} + \frac{J}{T} \EE_i \JJ_i^\fr \ ,
\eqend
by closing it with the constitutive relations in Eq.\,\eqref{constitutive.heatflux.current} and in Eq.\,\eqref{linear.eqs} or in Eq.\,\eqref{nonlin.eqs}. The governing equations will be rewritten as integral forms leading to a nonlinear and coupled weak form.\\

This nonlinear and coupled weak form will be solved numerically by using a staggered scheme since we solve electromagnetic potentials in the \textsc{Euler}ian frame, whereas the thermomechanical fields in the \textsc{Lagrange}an frame. Therefore, the scheme delivers a theoretically sound implementation principally suited for large deformation problems with fully coupled electromagnetic response of the material. Because of the staggered scheme introduced for solving fields in different frames, the accuracy of the numerical solution will be less than a monolithic solution procedure, as the trade-off for implementation flexibility.

\subsection{Variational formulation}

The whole computational domain is divided into non-overlapping finite elements with compact support. This discrete representation of the domain allows us to fulfill governing equations in finite elements. The equations are written as residuals and multiplied by a suitable test function in order to obtain scalar functions, which are then integrated over a suitable domain. This procedure is often called the variational formulation.\footnote{Technically, the terminology refers to obtaining this integral form by taking the first variation on an action. Since the outcome is identical, we use the same phrase.} As the suitable domain, often, the whole computational domain is selected; however, as the equations hold locally, it is also admissible to choose one finite element indicated by $\Om^\ele$. By summing over all elements composing the computational domain, we obtain the integral form to be solved. Within one finite element, the fields are $n$-times continuously differentiable and we will use standard \textsc{Lagrange} polynomial tetrahedral elements of order one such that their first derivative in space exists. Across the elements, the fields are continuous. We immediately replace all analytic functions with their corresponding discrete representations and omit a notational indication for the discrete functions in the following. \\

For the discretization in time, we use the backward \textsc{Euler} method since it is an implicit L-stable method. Let $t$ denote the current time being solved with timestep $\Delta t$ to advance the simulation from the previous time $t-\Delta t$. For any variable $K=\bar K(t)$ with a specified rate $\partial \bar K(t)/\partial t$, the implicit discretization in time uses a finite difference approximation between times $t-\Delta t$ and $t$ equated to the evaluation of the rate at time $t$. Simply by using a \textsc{Taylor} series around the current time and truncating after the linear term in $\Delta t$,
\begeq
K^0 = \bar K(t-\Delta t) = \bar K(t) - \Delta t \pd{\bar K(t)}{t} + O(\Delta t^2) \ , \\
\pd{\bar K(t)}{t} = \frac{K-K^0}{\Del t} \ ,
\eqend  
we immediately obtain the backward \textsc{Euler} method for evaluating the rate at the current time. The backward \textsc{Euler} method is unconditionally stable, but in our problem there is a limitation on the timestep size to guarantee convergence of the nonlinear problem that results to solve for $K(t)$. For second derivatives in time, we combine estimates of the first derivative at the current timestep and previous timestep to obtain
\begeq
\frac{\p^2 \bar K(t)}{\p t^2} = \frac{K-2K^0+K^{00}}{\Del t \Del t} \ ,
\eqend
where the superscript 00 denotes the value two timesteps before the current time, $K^{00}=\bar K(t-2\Del t)$. \\

We begin formulating the variational form to compute the electromagnetic fields in the \textsc{Euler}ian frame. The governing Eqs.\,\eqref{governing.em} as residuals read
\begeq \label{governing.em.computation}
\pd{q}{t} + \pd{J_i}{x_i} = 0 \comma
\pd{D_j}{t} - \epsilon_{jki} \pd{H_i}{x_k} +  J_j = 0 \ .
\eqend
The first scalar equation will deliver the scalar potential, $\phi$, and the second vector equations will serve for the vector potential, $\t A$. By following the aforementioned steps, we acquire the integral form,
\begeq
\int_{\Om^\ele} \bigg( \frac{q-q^0}{\Del t}  + \pd{J_i}{x_i} \bigg) \del\phi \d v = 0 \ .
\eqend
We multiply the form by $\Del t$ in order to bring it to the unit of energy. Moreover, we observe the derivative of the electric current, which includes $\t E$ depending on the derivative of $\phi$. Therefore, the unknown $\phi$ has to be twice differentiable in this form. The same condition holds for $q$ including the derivative of $\t D$ depending on $\t E$. This condition is weakened by integrating by parts, leading to the so-called weak form:
\begeq
\int_{\Om^\ele} \bigg( -(D_i-D_i^0) - \Del t J_i \bigg) \pd{\del\phi}{x_i} \d v + \int_{\p\Om^\ele} n_i (D_i-D_i^0 + \Del t J_i )  \del\phi\d a =0\ ,
\eqend
where $\t n$ is the unit normal on the element surface. If we sum up over all elements, which is called \textit{assembly} then the weak form becomes
\begeq
\sum_\text{ele.} \int_{\Om^\ele} \bigg( -(D_i-D_i^0) - \Del t J_i \bigg) \pd{\del\phi}{x_i} \d v + \sum_\text{sur.} \int_{\p\Om^\ele} \big\jl n_i (D_i-D_i^0 + \Del t J_i ) \del\phi \big\jr  \d a 
+\\
+ \sum_\text{outer} \int_{\p\Om^\ele} n_i (D_i-D_i^0 + \Del t J_i )  \del\phi\d a =0\ .
\eqend
Three distinct summations are applied: summation over the elements, summation over the inner surfaces with two adjacent elements, and summation over the outer surfaces being on the computational domain boundary. It is necessary to approximate infinite or far-field boundary conditions for the electromagnetic fields. The computational domain is extended a finite distance far away from the solid body where electromagnetic potentials vanish; this is often implemented by placing the solid body inside of a very large sphere in the meshing software. We set $\phi$ at the computational domain via \textsc{Dirichlet} boundary condition such that its test function on the outer $\p\Om$ vanishes. For the term in jump brackets we use Eqs.\,\eqref{sing.em1} as follows
\begeq
\big\jl n_i (D_i-D_i^0 + \Del t J_i ) \del\phi \big\jr 
= \Big\jl n_i \Big( \Del t J_i^\fr + P_i-P_i^0 + \Del t \epsilon_{ijk} \pd{\MM_k}{x_j} \Big) \del\phi \Big\jr
= \Big\jl n_i \Del t \Big(  J_i^\fr + \epsilon_{ijk} \pd{\MM_k}{x_j} \Big) \del\phi \Big\jr \ .
\eqend
Finally, we obtain the weak form for computing the electric potential in the \textsc{Euler}ian frame,
\begeq \label{weak.1}
\text{F}_\phi = \sum_\text{ele.} \int_{\Om^\ele} \bigg( -(D_i-D_i^0) - \Del t J_i \bigg) \pd{\del\phi}{x_i} \d v 
+ \sum_\text{sur.} \int_{\p\Om^\ele} \Big\jl n_i \Del t \Big(  J_i^\fr + \epsilon_{ijk} \pd{\MM_k}{x_j} \Big) \del\phi \Big\jr  \d a
\ .
\eqend
There are different possible approaches to obtain the weak form for the magnetic potential, $\t A$. We follow the approach suggested in \cite{027} that leads to a robust computational method. First, we rewrite Eq.\,\eqref{governing.em.computation}$_2$ by inserting the \textsc{Maxwell--Lorentz} aether relations and then the electromagnetic potentials from Eq.\,\eqref{em.ansatz} as follows:
\begeq
\pd{\eps_0 E_j}{t} - \epsilon_{jki} \pd{}{x_k} \Big( \frac1{\mu_0} B_i \Big) + J_j = 0 
\ , \\
-\eps_0 \pd{^2 \phi}{t\p x_j} - \eps_0 \pd{^2 A_j}{t^2} - \frac1{\mu_0} \epsilon_{jki} \epsilon_{ilm} \pd{^2 A_m}{x_k \p x_l} +  J_j = 0 
\ , \\
-\pd{}{x_j} \Big( \eps_0 \pd{\phi}{t} + \frac1{\mu_0} \pd{A_k}{x_k} \Big) - \eps_0 \pd{^2 A_j}{t^2} + \frac1{\mu_0} \pd{^2 A_j}{x_k \p x_k} +  J_j = 0 \ ,
\eqend
using the identity $\epsilon_{jki} \epsilon_{ilm}=\delta_{jl}\delta_{km}-\delta_{jm}\delta_{kl}$. The first term in brackets vanishes as a consequence of \textsc{Lorenz}'s gauge in Eq.\,\eqref{lorenz}. Then the variational formulation, \begeq
\int_{\Om^\ele} \bigg( - \eps_0 \frac{A_j-2A_j^0+A_j^{00}}{\Del t \Del t}  + \frac1{\mu_0} \pd{^2 A_j}{x_k \p x_k} + J_j \bigg) \del A_j  \d v = 0 \ , 
\eqend
delivers
\begeq
\int_{\Om^\ele} \bigg( - \eps_0 \frac{A_j-2A_j^0+A_j^{00}}{\Del t \Del t} \del A_j - \frac1{\mu_0} \pd{A_j}{x_k } \pd{\del A_j}{x_k} + J_j^\fr \del A_j + \frac{P_j-P_j^0}{\Del t} \del A_j 
- \\
-\epsilon_{jki} \MM_i \pd{\del A_j}{x_k} \bigg)   \d v 
+ \int_{\p\Om^\ele} \bigg( \frac1{\mu_0} \pd{A_j}{x_k } + \epsilon_{jki} \MM_i \bigg) \del A_j n_k \d a =0\ ,
\eqend
after an integration by parts on the terms already including a derivative. The integral form is in the unit of energy. The assembly generates a jump term that vanishes for continuous magnetic potential and by using Eq.\,\eqref{sing.em2}. Furthermore, we set the magnetic potential zero at the computational domain boundary. Hence, the weak form for the magnetic potential reads in the \textsc{Euler}ian frame,
\begeq \label{weak.2}
\text{F}_{\t A} = \sum_\text{ele.} \int_{\Om^\ele} \bigg( - \eps_0 \frac{A_j-2A_j^0+A_j^{00}}{\Del t \Del t} \del A_j - \frac1{\mu_0} \pd{A_j}{x_k } \pd{\del A_j}{x_k} + J_j^\fr \del A_j + \frac{P_j-P_j^0}{\Del t} \del A_j  -\epsilon_{jki} \MM_i \pd{\del A_j}{x_k} \bigg)   \d v 
\ .
\eqend \\

In the case of thermomechanics, we solve $T$ and $u_i$ in the \textsc{Lagrange}an frame. After discretizing the variational form in time and integrating by parts, the balance of linear momentum reads for a finite element
\begeq
\int_{\Bo^\ele} \bigg( \rho_0 \frac{u_i - 2u_i^0 + u_i^{00}}{\Del t \Del t} \del u_i + J (\t F^{-1})_{kj} \sigma_{ji} \pd{\del u_i}{X_k} \del u_i - \rho_0 f_i \del u_i - J\elF_i \del u_i \bigg) \d V
- \\
- \int_{\p\Bo^\ele} J (\t F^{-1})_{kj} \sigma_{ji} \del u_i N_k \d A =0\ ,
\eqend
where we distinguish between the infinitesimal elements, plane normals, as well as domains in the \textsc{Lagrange}an and \textsc{Euler}ian frames for the sake of clarity. The latter integral form is in the unit of energy. The assembly results in a vanishing jump term in connection with Eq.\,\eqref{sing.tm1}. On the boundaries, where the boundary for the thermomechanics means the interface to the surrounding air, we either set the displacements by using a \textsc{Dirichlet} condition or set the applied traction force per area $\hat t_i=N_k J (\t F^{-1})_{kj}\sigma_{ji}$ by using a \textsc{Neumann} condition. The weak form for computing the displacement reads
\begeq \label{weak.3}
\text{F}_{\t u} = \sum_\text{ele.} \int_{\Bo^\ele} \bigg( \rho_0 \frac{u_i - 2u_i^0 + u_i^{00}}{\Del t \Del t} \del u_i  + J (\t F^{-1})_{kj} \sigma_{ji} \pd{\del u_i}{X_k} - \rho_0 f_i \del u_i - J \elF_i \del u_i \bigg) \d V 
+\\
+ \sum_\text{outer} \int_{\p\Bo^\ele} \hat t_i \del u_i \d A \ .
\eqend
Analogously, to compute the temperature, we obtain the weak form by using the time discretization, applying the variational formulation, multiplying by $\Del t$ in order to bring it to the unit of energy, and integrating by parts to obtain
\begeq
\int_{\Bo^\ele} \bigg( \rho_0 (\eta - \eta^0) \del T - \Del t \frac{Q_j}{T} \pd{\del T}{X_j} - \Del t \rho_0 \frac{r}{T} \del T + \Del t \frac{Q_j}{T^2} \pd{T}{X_j} \del T - \Del t \frac{J}{T} \EE_i \JJ_i^\fr \del T \bigg) \d V
+ \\
+ \int_{\p\Bo^\ele} \Del t \frac{Q_j}{T} \del T N_j \d A =0\ .
\eqend
The assembly results in a jump term that vanishes between the elements by means of Eqs.\,\eqref{sing.tm2}, \eqref{int.energy.mat.frame}$_2$. On the interface to the surrounding air, we model this jump as follows
\begeq
\Big\jl Q_j N_j \Big\jr = h (T - T_\Ref) \ ,
\eqend
with the convective heat transfer coefficient $h$ furnishing an exchange between the continuum body and environment depending on the velocity of the surrounding fluid. This approximation is necessary since we skip to compute the velocity of the surrounding air directly. Analogously, on the computational boundary we set $T=T_\Ref$ as \textsc{Dirichlet} boundaries. The weak form to compute $T$ in the \textsc{Lagrange}an frame reads
\begeq \label{weak.4}
\text{F}_{T} = \sum_\text{ele.} \int_{\Bo^\ele} \bigg( \rho_0 (\eta - \eta^0) \del T - \Del t \frac{Q_j}{T} \pd{\del T}{X_j} - \Del t \rho_0 \frac{r}{T} \del T + \Del t \frac{Q_j}{T^2} \pd{T}{X_j} \del T - \Del t \frac{J}{T} \EE_i \JJ_i^\fr \del T \bigg) \d V 
+ \\
+ \sum_\text{outer} \int_{\p\Bo^\ele} \Del t  \frac{h(T-T_\Ref)}{T} \del T \d A \ .
\eqend
We recall that we solve for the same time instant $\text{F}_\phi+\text{F}_{\t A}$ in the \textsc{Euler}ian frame and $\text{F}_{\t u}+\text{F}_{T}$ in the \textsc{Lagrange}an frame.

\subsection{Mesh morphing}

The weak form $\text{F}_\phi+\text{F}_{\t A}$ in Eqs.\,\eqref{weak.1} and  \eqref{weak.2} is solved in the \textsc{Euler}ian frame for the whole computational domain, i.e., continuum body embedded in air (or vacuum). In order to solve the weak forms in the \textsc{Euler}ian frame, we need to move the continuum body to its current placement. The idea is similar to the same technique used in fluid-structure interaction; however, the electromagnetic fields are also described within the continuum body. The weak form $\text{F}_{\t u}+\text{F}_T$  in Eqs.\,\eqref{weak.3}, \eqref{weak.4} needs to be solved in the \textsc{Lagrange}an frame only within the continuum body and not in the surrounding air. The motion of the continuum body within the computational domain connects these two frames such that we move the mesh by the displacement of the continuum body. This particular choice is justified by Eq.\,\eqref{sing.boundary.vel}. For the surrounding air, we spare solving the displacement such that we morph the surrounding domain in a particular way explained below in order to maintain the mesh quality. \\

The mesh for the problem is constructed on the entire domain $\Omega$, enveloping the solid body and a sufficiently large air region that extends into the far-field. The region corresponding to the body is marked in the meshing software. The FEM simulation extracts the elements corresponding to solid materials to create a second ``submesh'' object for the domain $\mathcal{B}_0$ in the reference state; the elements and nodes for the mesh of  $\mathcal{B}_0$ are contained in the mesh of $\Omega$, but reordering of nodes occurs to assemble the variational forms of $\text{F}_u+\text{F}_T$ on only this submesh. The nodal coordinates in the mesh correspond to the initial placement $\text{X}$ of the material points. \\

\begin{figure}
\centering
\includegraphics[width=0.85\linewidth]{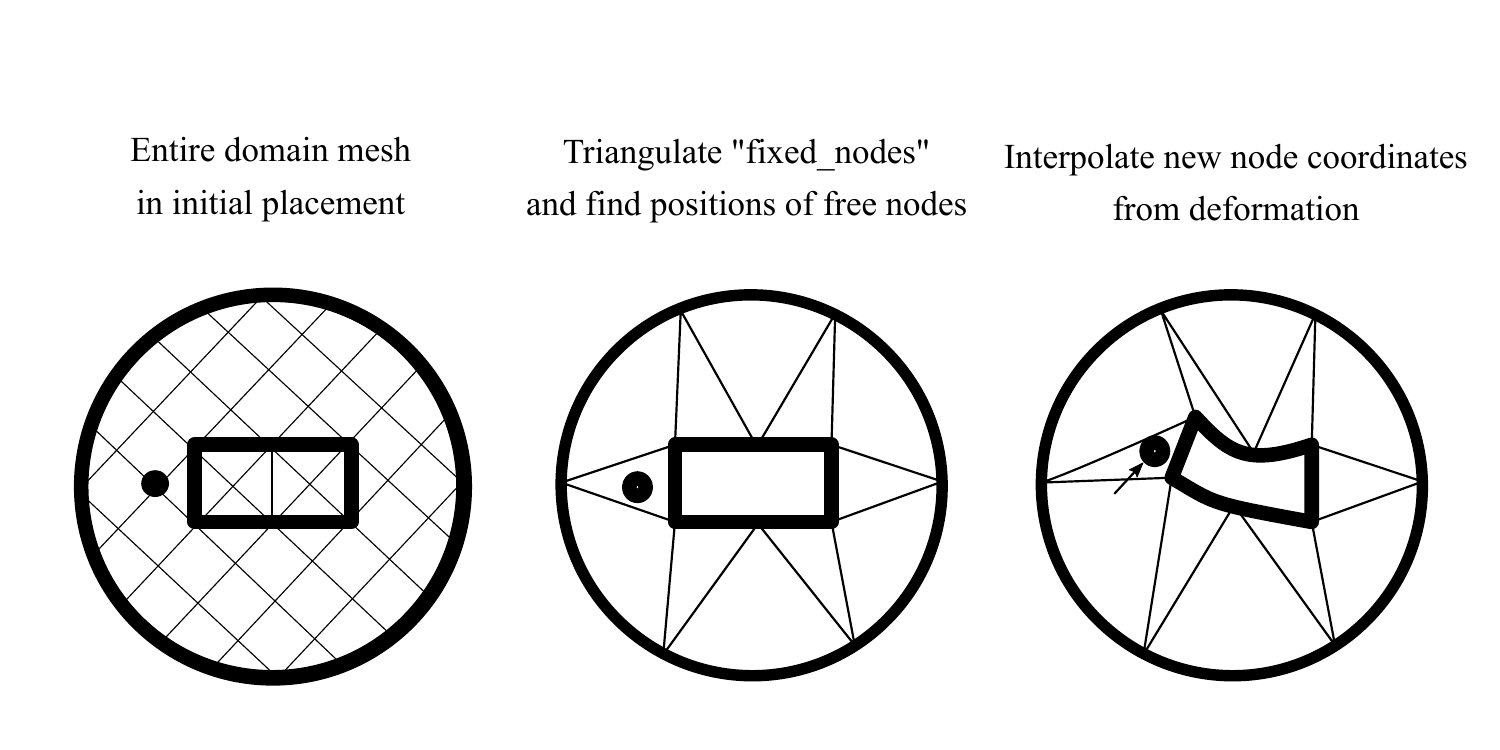}
 \caption{Illustration of the mesh morphing procedure. The original mesh is separated into nodes on the external boundary and solid body, the ``fixed\_nodes,'' and those in the interior of the air domain. In the diagram, consider one node in the mesh covering the air. The node's position is determined on the triangulation of the ``fixed\_nodes.'' During the simulation, the solid body deforms, and the spanning triangulation is used to ``drag'' the air node to a new location.}
 \label{fig:morphing}
\end{figure}

The procedure is illustrated in Figure \ref{fig:morphing} and listed in Algorithm \ref{alg:morphing}. The part of the mesh excluding the solid body, $\Omega \setminus \mathcal{B}_0$, corresponds to the electromagnetic-only domain. The mesh morphing algorithm will determine new positions to these nodes on the interior of the domain in response to the motion of the body. The indices of the nodes on the external boundary and inside of the body $X[i] \in \partial \Omega \cup \mathcal{B}_0$ are labeled as ``fixed\_nodes''. The positions of the fixed nodes, $X_f=X[fixed\_nodes]$, are used to interpolate the positions of the rest of the nodes. A Delaunay triangulation (or tetrahedralization), $tris$, is constructed from the positions of fixed nodes. The list of points of $\textit{fix\_nodes}$ should contain a convex hull enclosing the domain, otherwise the interpolation is ill-defined. This condition is met by making sure that the boundary of the far-field mesh $\partial \Omega$ is included in the list. \\

During the solution procedure, the solid body will move to a deformed state, $X_f'$. The nodes on the exterior boundary $\partial \Omega$ remain at their orignal position. The remaining nodes in the mesh, those in the surrounding air, are ``glued'' to the triangulation. For node, the barycentric coordinates in the initial configuration are determined. The new positions of the air nodes are calculated by using the shape functions of the triangulation interpolating the new positions, $X' = \sum_i b_i X'[tri[i]]$. The element connectivity does not change. The mesh velocity at each node is then calculated from this update step given the time step by $ w = (X'-X)/\Delta t $. \\

The quality of the triangulation does not matter as it is only used to move nodes; the initial mesh connectivity is still used to calculate the finite element fields.
A mesh-smoothing step could be applied after the interpolation of new coordinates to improve the morphed quality. This step was not needed in the simulations in the next section. The elements in the triangulation does not need to span the body surface and exterior domain boundary; the body can have ``cavities'' with air inside of it as well having, for example, a capacitor or C-circuit geometry. \\

The Delaunay triangulation is computed using the Scipy module, which provides the barycentric coordinate transformations and fast point-in-triangle detection \cite{oliphant2007}, \cite{millman2011}. The barycentric coordinate transformations are a by-product of the algorithm and fortuitously correspond to the shape functions needed for the interpolation of new nodal coordinates. \\

\begin{algorithm}[htb]
  \caption{Calculate Mesh morphing procedure. Note, the variable $tris$ is a list of tetrahedra in 3D.}
  \label{alg:morphing}
  \begin{algorithmic}
    \REQUIRE Nodes $X$, indices $\textit{fixed\_nodes}$
    \STATE $n_g \leftarrow $ Geometric dimension (2 or 3)
    \STATE $X_f \leftarrow X[\textit{fixed\_nodes}]$
    \STATE $tris \leftarrow \mathrm{Delaunay}(X_f)$
    \STATE Preallocate $X'$
    \FOR{$y \in X$ \AND $y \notin X_f$}
    \IF{$y \notin X_f$}
    \STATE Find $t \in tris \ s.t.\  y\in t$ using search tree.
    \STATE $B \leftarrow $ barycentric transformation matrix of $t$
    \STATE $s[0:n_g-2] \leftarrow B \cdot y$
    \STATE $s[n-1] \leftarrow 1-sum(s[0:n_g-2])$
    \STATE $y'[i] \leftarrow s[j]*X_f[t[j]][i] $ for $j=0...n_g$ for $i=0...n_g-1$
    \STATE $X'[end] \leftarrow y'$
    \ELSE
    \STATE $X'[end] \leftarrow y$
    \ENDIF
    \ENDFOR
    \RETURN $X'$
  \end{algorithmic}
\end{algorithm}

\subsection{Coupled time stepping algorithm}

The thermomechanical fields $\t u$ and $T$ are defined on a submesh of the computational domain corresponding the material body. This submesh is defined as $\textit{matmesh}$ The electromagnetic potentials $\phi$ and $\t A$ are defined on the mesh of the whole computational domain. They are solved using a sequential (staggered) scheme in each time step. Then, the finite element coefficients from the electromagnetic finite element functions are mapped onto the equivalent coefficients on the submesh of the domain occupied by the continuum body. Then, the thermomechanical problem is solved for $\t u$ and $T$ at the next time-step. The displacement $\t u$ is used to morph the mesh and then the thermomechanical fields are mapped onto the morphed mesh of the whole domain. The procedure of the time stepping algorithm is listed in Algorithm \ref{alg:main}.

\begin{algorithm} [htb]
  \caption{\label{alg:main} Main Simulation Procedure}
  \begin{algorithmic}
    \STATE $mesh \leftarrow$ File
    \STATE $matmesh \leftarrow\subset mesh$ where $mesh$ is marked as material. 
    \STATE $fix\_nodes \leftarrow vertex\_indices[matmesh \cup \partial mesh]$
    \STATE $u^{EM},A^{EM}\leftarrow$ Thermomechanical and electromagnetic functions on $mesh$
    \STATE $u^{TM},A^{TM}\leftarrow$ Thermomechanical and electromagnetic functions on $matmesh$
    
    \WHILE{$t < t_{max}$}
    \STATE Pull solution coefficients onto $matmesh$: $A^{TM}.\textit{coeffs}[:]\leftarrow A^{EM}.\textit{coeffs}[indices]$
    \STATE Solve $F^{TM}(u^{TM} ; A^{TM})=0$ for $u^{TM}(t+\Delta t)$
    \STATE Push solution coefficients onto $mesh$: $u^{EM}.\textit{coeffs}[indices]\leftarrow u^{TM}.\textit{coeffs}[:]$
    \STATE $mesh.X\leftarrow$ morph$(mesh.X+u^{EM},fix\_nodes)$
    \STATE Solve $F^{EM}(A^{EM} ; u^{EM})=0$ for $A^{EM}(t+\Delta t)$

    \STATE $t\leftarrow t+\Delta t$
    \STATE Advance history: $A(t+\Delta t)\rightarrow A(t)\rightarrow A(t-\Delta t)$
    \ENDWHILE
  \end{algorithmic}
\end{algorithm}

\subsection{Implementation Details}

The entire algorithm is implemented in Python using the open-source packages developed under the FEniCS project, see \cite{dolfin,fenics,fenics_book}. The mesh morphing algorithm, as well as other utility functions, is included for use in different applications in the library afqsfenicsutils.\footnote{The git repository is located at \url{https://bitbucket.org/afqueiruga/afqsfenicsutil}} The FEniCS implementation for the variational equations and coupling system is available at the repository located at \url{https://github.com/afqueiruga/EMSI-2018}. (This repository will automatically download the independent libraries as a git submodule.) All codes are released under the GNU Public license as in \cite{gnupublic}. 

\section{Examples}
\label{sec:examples}
We illustrate the functionality and versatility of the framework based upon the presented theoretical formulation and numerical algorithm by applying it to three engineering applications. We have chosen the examples where the interaction of thermomechanics with electromagnetism generates new design opportunities necessitating robust and accurate computation of such coupled and nonlinear multiphysics problems by using the simulation strategy developed herein. \\

Smart structure applications use materials with integrated sensor and actuator functionalities. There are only a few known natural materials presenting electromagnetic and mechanical coupling inherently. Therefore, functionalized materials are synthesized by combining materials with different abilities. In various branches of industry, these types of functionalized materials are applied in applications such health monitoring; shape, temperature, or vibration control; or energy harvesting. For examples by using a piezoelectric material, we refer to \cite{yamada1988}, \cite{losinski1999}, \cite{park2003}, \cite{sodano2005}, \cite{kim2006}, \cite{kovalovs2007}, \cite{lanza2007}, \cite{brunner2009}, \cite{yang2009}, \cite{paradies2009}, \cite{van2011}, \cite{tanida2013}, \cite{pagel2013}. In \cite{ginder1999}, \cite{frommberger2003}, \cite{ausanio2005}, \cite{bienkowski2010}, \cite{grimes2011}, \cite{li2013}, \cite{guo2017}, a magnetoelastic material is used for different applications. Thermoelectric coupling is demonstrated in \cite{sauciuc2006}, \cite{zhao2014}, \cite{he2015}.
 We present three examples that are representative of new industry applications: a piezoelectric fan, a magnetorheological elastomer, and a thermoelectric cooler.
\subsection{Piezoelectric fan}

\begin{figure} 
\centering
\includegraphics[width=0.6\linewidth]{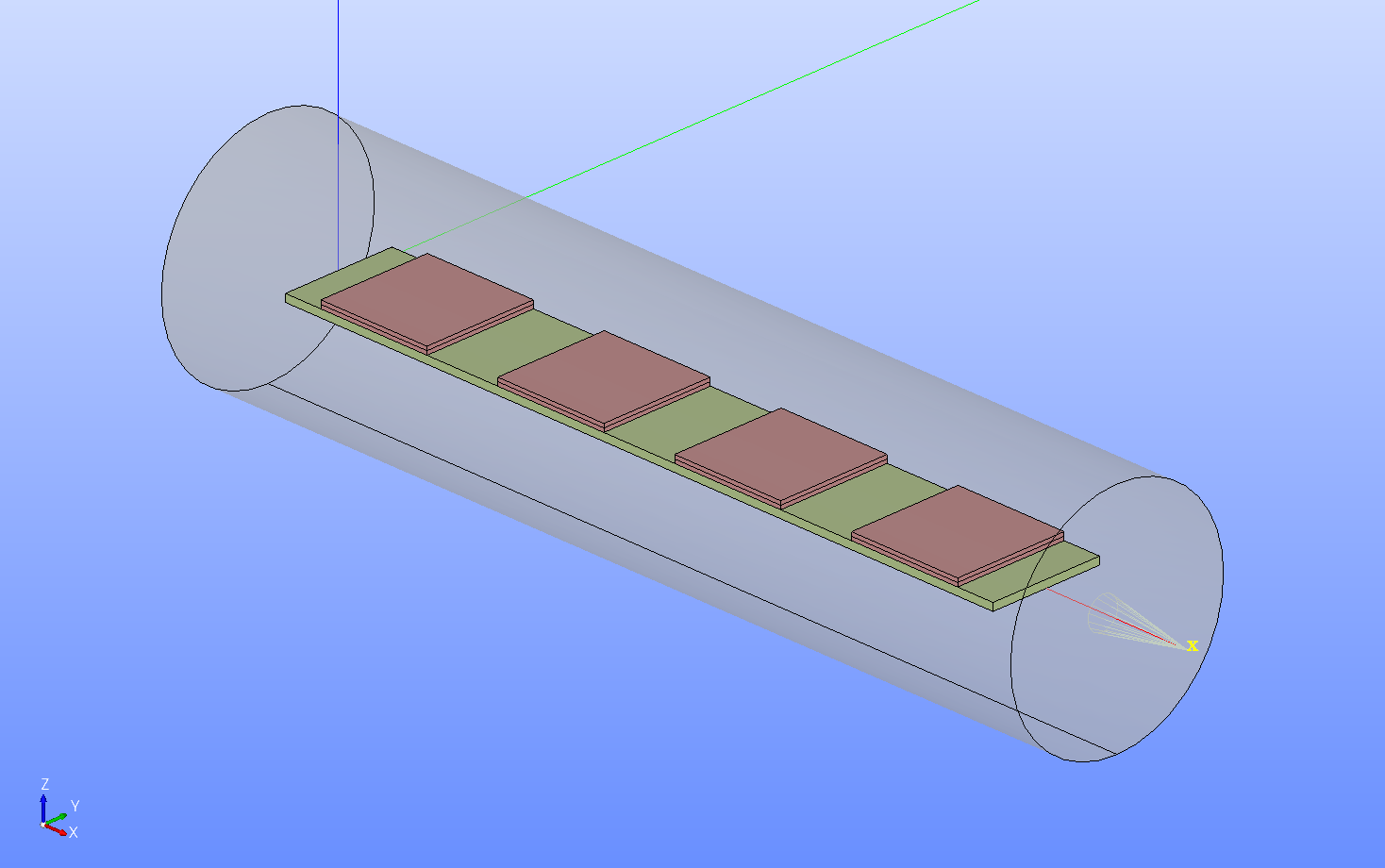}
 \caption{CAD model of the piezoelectric fan: on the epoxy beam (green), four piezoelectric patches (red) are attached comprising the continuum body $\Bo$ embedded in air (gray, transparent).}
 \label{fig:piezo.1}
\end{figure}

\begin{figure} 
\centering
\includegraphics[width=0.49\linewidth]{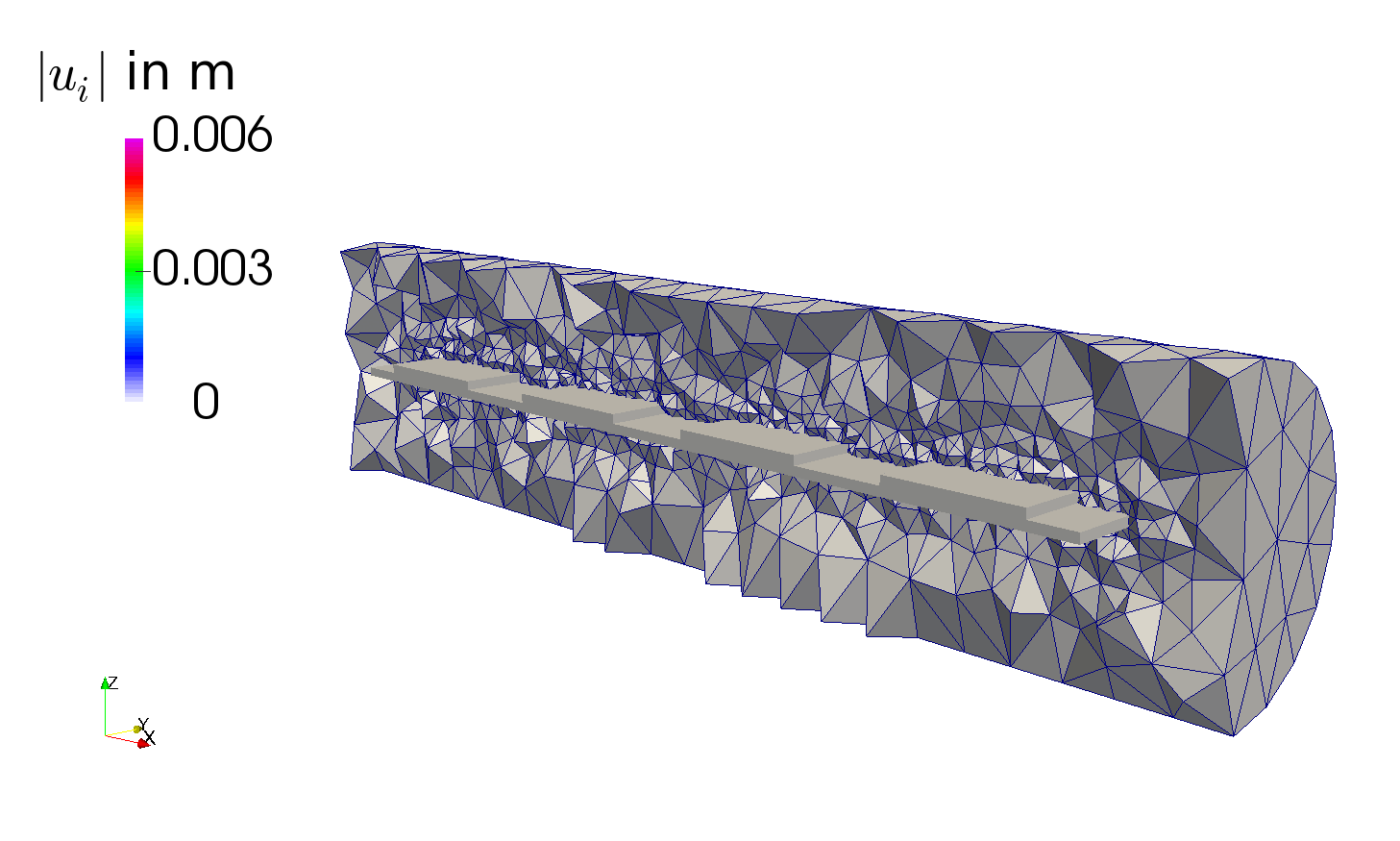}
\includegraphics[width=0.49\linewidth]{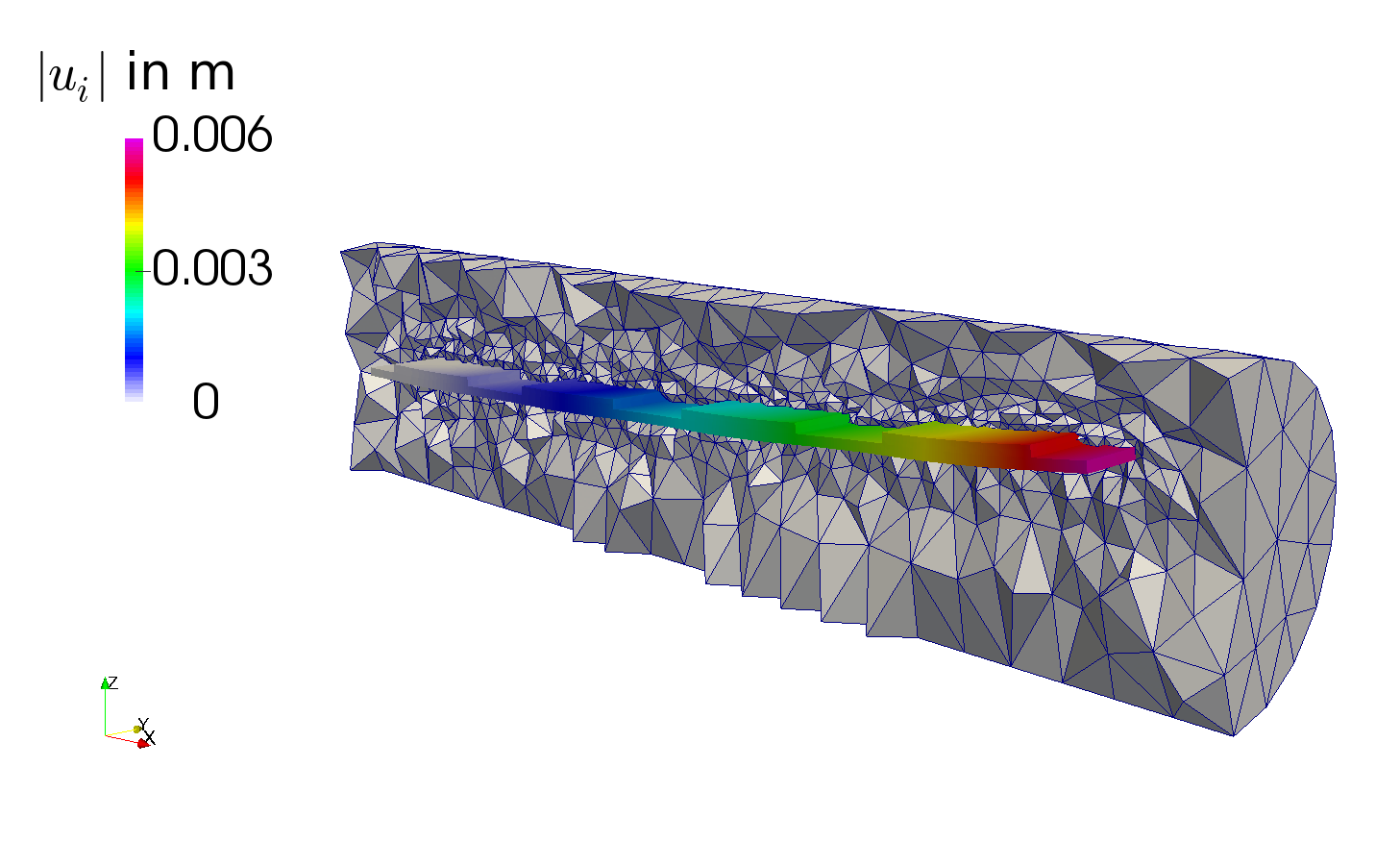}
 \caption{Configuration of the beam bend by the piezoelectric patches. The continuum body is colored by the magnitude of the displacement as well as morphed by the displacement {\em without} scaling. The air mesh is not colored and has a crinkle cut to reveal the fan and illustrate the morphing of the elements.}
 \label{fig:piezo.2}
\end{figure}

\begin{figure}
\centering
\includegraphics[width=0.49\linewidth]{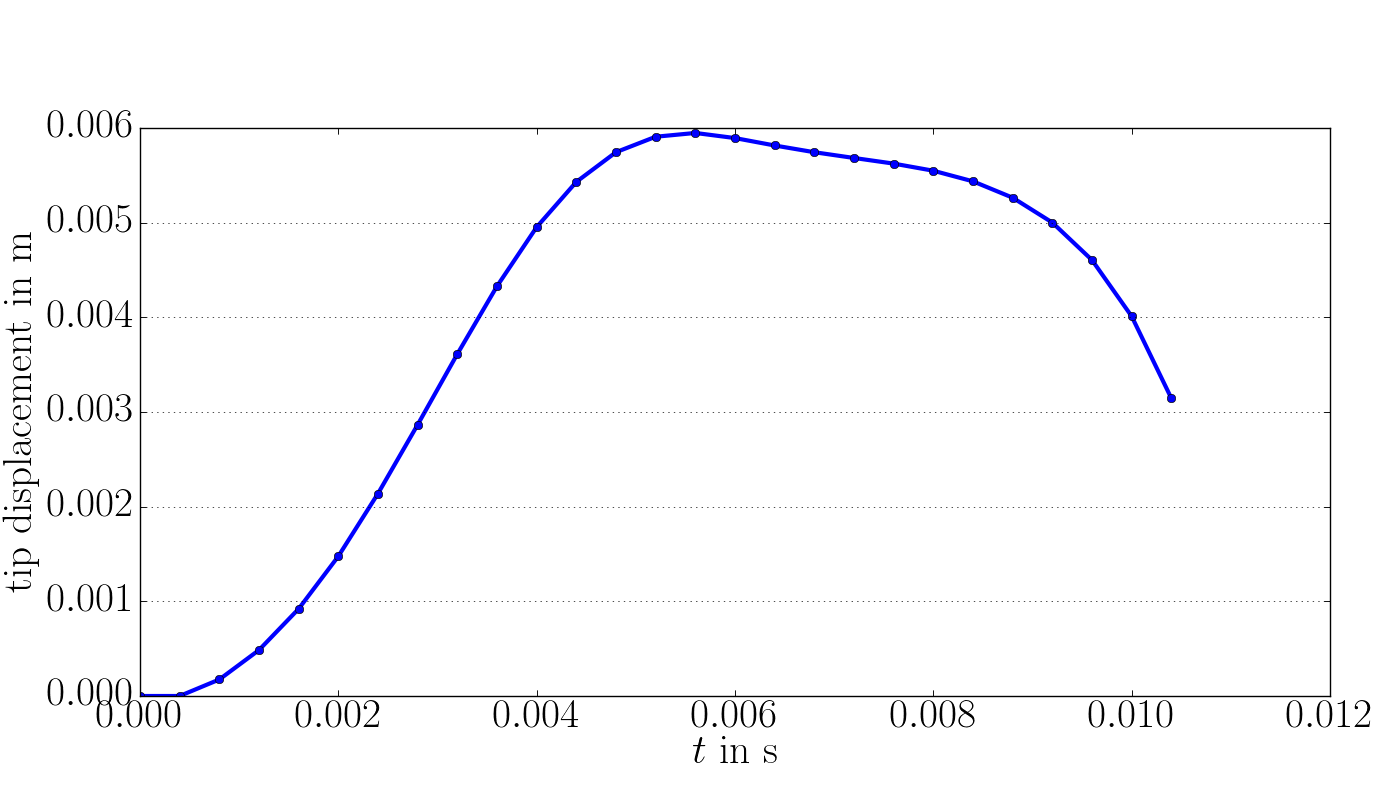}
\includegraphics[width=0.49\linewidth]{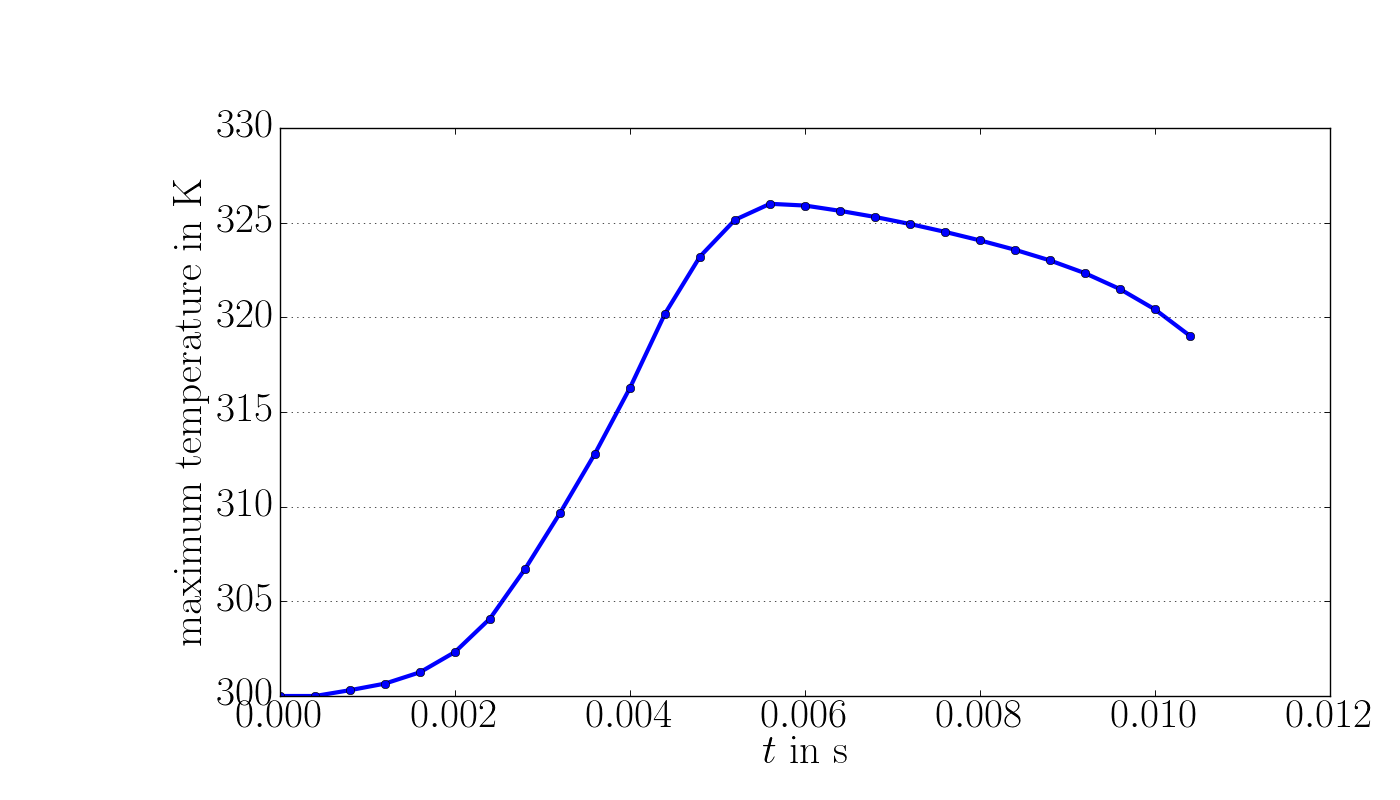}
 \caption{Displacement of the tip of the device (left) and the maximum temperature in the device (right) over the course of the simulation.}
\label{fig:piezo.3}
\end{figure}

A thin beam of 100\,mm\,$\times$\,15\,mm\,$\times$\,1\,mm is modeled out of epoxy with four piezoelectric patches. Each patch is of two layers connected in parallel. In practice, many layers are used to increase the effect, herein we present a simplified computation with the geometry shown in Fig.\,\ref{fig:piezo.1}. The continuum body, $\Bo$, is embedded in a cylindrical domain, modeling surrounding air, with far-away boundaries where the electromagnetic potentials vanish.
Piezoelectric patches are made of 2 layers each of $0.5$\,mm thickness. They are poled along $z$-direction. For the piezoceramic and epoxy, we use the stiffness matrix $C_{IJ}$ in \textsc{Voigt}'s notation
\begeq
C_{IJ}=
\begin{pmatrix}
C_{1111} & C_{1122} & C_{1133} & C_{1123} & C_{1113} & C_{1112} \\
C_{2211} & C_{2222} & C_{2233} & C_{2223} & C_{2213} & C_{2212} \\
C_{3311} & C_{3322} & C_{3333} & C_{3323} & C_{3313} & C_{3312} \\
C_{2311} & C_{2322} & C_{2333} & C_{2323} & C_{2313} & C_{2312} \\
C_{1311} & C_{1322} & C_{1333} & C_{1323} & C_{1313} & C_{1312} \\
C_{1211} & C_{1222} & C_{1233} & C_{1223} & C_{1213} & C_{1212}
\end{pmatrix} \ .
\eqend
Epoxy is an amorph material having translational and rotational symmetry such that the stiffness matrix is isotropic,
\begeq
C_{IJ}=
\begin{pmatrix}
\lambda+2\mu & \lambda & \lambda & 0 & 0 & 0 \\
\lambda & \lambda+2\mu & \lambda & 0 & 0 & 0 \\
\lambda & \lambda & \lambda+2\mu & 0 & 0 & 0 \\
0 & 0 & 0 & \mu & 0 & 0 \\
0 & 0 & 0 & 0 & \mu & 0 \\
0 & 0 & 0 & 0 & 0 & \mu
\end{pmatrix} \comma
\lambda = \frac{(E-2G)G}{3G-E}
\comma
\mu = G \ ,
\eqend
with \textsc{Young}'s modulus $E$ and shear modulus $G$. As piezoceramic we use PZT-5H poled along $z=x_3$. For this anisotropic PZT-5H, the compliance matrix 
\begeq
S_{IJ}= \begin{pmatrix} S_{11} & -\nu S_{11} & -\nu S_{11} & 0 & 0 & 0 \\ -\nu S_{11} & S_{11} & -\nu S_{11} & 0 & 0 & 0 \\ -\nu S_{11} & -\nu S_{11} & S_{33} & 0 & 0 & 0 \\ 0 & 0 & 0 & (1+\nu)S_{11} & 0 & 0 \\ 0 & 0 & 0 & 0 & (1+\nu)S_{11} & 0 \\ 0 & 0 & 0 & 0 & 0 & (1+\nu)S_{11} \end{pmatrix} \ ,
\eqend
is used to obtain the stiffness matrix $C_{IJ}=(S_{JI})^{-1}$. The piezoelectric constants, $\tilde d_{iJ}$, read
\begeq
\tilde d_{i J} = 
\begin{pmatrix} \tilde d_{111} & \tilde d_{122} & \tilde d_{133} & \tilde d_{123} & \tilde d_{131} & \tilde d_{112} \\ 
\tilde d_{211} & \tilde d_{222} & \tilde d_{233} & \tilde d_{223} & \tilde d_{231} & \tilde d_{212} \\ 
\tilde d_{311} & \tilde d_{322} & \tilde d_{333} & \tilde d_{323} & \tilde d_{331} & \tilde d_{312} \end{pmatrix}
=
\begin{pmatrix} 0&0&0&0&\tilde d_{15}&0 \\ 0&0&0&\tilde d_{15}&0&0 \\ \tilde d_{31}&\tilde d_{31}&\tilde d_{33}&0&0&0 \end{pmatrix} \ ,
\eqend
where \textsc{Voigt}'s notation is applied on the last two indices (mapping to multiplication by the displacement gradient in \textsc{Voigt}'s notation). The susceptibility is given by the relative permittivity values by 
\begeq
\chi^\el_{ij} = \begin{pmatrix} \bar\eps^\el_{11} & 0 & 0 \\ 0 & \bar\eps^\el_{11} & 0 \\ 0 & 0 & \bar\eps^\el_{33} \end{pmatrix} - \delta_{ij} \ .
\eqend
We assume that the material has no piezomagnetic and magnetoelectric coupling, i.e., $\tilde S_{ijk}=0$ and $\tilde R_{ij}=0$, respectively. We compile all necessary material parameters in Table~\ref{tab:mat.param}. Thermoelectric constant and electric conductivity is set to zero for the beam and patches.
\begin{table}[!hbt]
\caption{Material constants used in the simulation for the epoxy material, PZT-5H as the piezoceramic, and the surrounding air.}
\centering
\renewcommand{\arraystretch}{1.2}
\begin{tabular}{p{5.5cm}|c|c|c|c|}
\toprule
& & Epoxy & PZT-5H & Air \\
\midrule
Mass density & $\rho$ in kg/m$^3$ & 2500 & 7500 &  \\
\midrule
\multirow{6}{*}{Compliance} & $S_{11}$ in m$^2$/N &  & $16.5\cdot 10^{-12}$ &  \\
                            & $S_{12}$ in m$^2$/N &  & $-4.78\cdot 10^{-12}$ &  \\
                            & $S_{13}$ in m$^2$/N &  & $-8.45\cdot 10^{-12}$ &  \\
                            & $S_{33}$ in m$^2$/N &  & $20.7\cdot 10^{-12}$ &  \\
                            & $S_{44}$ in m$^2$/N &  & $43.5\cdot 10^{-12}$ &  \\
                            & $S_{66}$ in m$^2$/N &  & $42.6\cdot 10^{-12}$ &  \\
\midrule
\textsc{Young}'s modulus & $E$ in N/m$^2$ & $30\cdot 10^9$ &  &  \\
\midrule
\textsc{Poisson}'s ratio & $\nu$ & $0.4$ &  &  \\
\midrule
\multirow{3}{*}{Piezoelectric constants} & $\tilde d_{33}$ in m/V & 0 & $585\cdot 10^{-12}$ & \\
                                        & $\tilde d_{31}$ in m/V & 0 & $-265\cdot 10^{-12}$ & \\
                                        & $\tilde d_{15}$ in m/V & 0 & $730\cdot 10^{-12}$ & \\
\midrule
\multirow{2}{*}{Dielectric constants} & $\bar\eps^\el_{33}$ & 1 & 3400 & 1\\
                                      & $\bar\eps^\el_{11}$ & 1 & 3130 & 1\\
\midrule
Specific heat capacity & $c$ in J/(kg\,K) & 800 & 350 & \\
\midrule
\multirow{2}{*}{Coefficients of thermal expansion} & $\alpha_{33}$ in K$^{-1}$ & $15\cdot 10^{-6}$ & $-4\cdot 10^{-6}$ &  \\
                                      & $\alpha_{11}$ in K$^{-1}$ & $15\cdot 10^{-6}$ & $6\cdot 10^{-6}$ &  \\
\midrule
Thermal conductivity & $\kappa$ in W/(m\,K) & 1.3 & 1.1 & \\
\bottomrule
\end{tabular}
\label{tab:mat.param}
\end{table}
We apply a sinusoidal electric potential difference on the piezoelectric patches by grounding the bottom and upper faces and changing the middle surface in time. Along the $z$-axis an electric field emerge that leads to a contraction along $x$ as well as $y$-axis because of $\tilde d_{31}$. Since the potential difference from the middle to the top layer and from the middle to the bottom layer produces in electric fields that are opposed to  other in the each layer, one layer stretches when the other layer contracts. The bending in each patch bends the entire beam as shown in Fig.\,\ref{fig:piezo.2}. We have applied a relatively big potential difference (amplitude) of 50\,kV in order to generate a big deformation by using only 2 layers of patches.
The displacement of the tip and the maximum temperature in the device over the course of the simulation is plotted in Fig.\,\ref{fig:piezo.3}. Effected by the exaggerated potential difference, a significant temperature change occurs because of the electric field jump on the middle layer is generated as presented. Further engineering on this type of device would be needed to reduce the required potential difference and resultant heat production, which is possible due to the fully coupled simulation demonstrated.

\subsection{Magnetorheological elastomer}

The deformation and magnetic field coupling is often called magnetostriction; but it is insignificant in natural materials. By designing a functionalized material, this behavior is used extensively for smart structures. Consider an elastomer filled with iron spherical particles with sizes on the order of micrometers. To model this material at the macroscopic scale, we homogenize the material into a magnetorheological elastomer. The thermomechanical behavior of the composite will be primarily representative of the elastomer matrix, with additional electromagnetic properties due to the iron additives. Because the iron particles are spherical, an elastomer with an amorph structure will remain isotropic if no external magnetic field was applied during the curing, see \cite{li2013}. This crystalline structure with inversion symmetry prohibits any piezoelectric effects, $\tilde T_{ijk}=0$. We assume that the magnetoelectric coupling vanishes, $\tilde R_{ij}=0$. A piezoelectric effect is possible depending on the crosslinking of the polymer chains in the elastomer. A magnetoelectric coupling is also expected to arise as a consequence of this effect. The computational framework could include this effect with the necessary material constants, but it was neglected for this simulation. \\

The functionalized material considered is taken to be a silicone gel TSE2062 filled with carbonyl-iron particles. By assuming an equal and distinct distribution and successful curing, the elastomer has the thermomechanical properties of the silicon. This approximation depends on the relative amount of the iron particles used in the manufacturing. Increasing the amount leads to agglomerated particles building ``bridges'' between the distinct iron particles such that the thermomechanical characteristics of the composite material change dramatically. For an accurate treatment we refer to \cite{zohdi2008} and \cite{zohdi2012}. The material properties of the composite material---the particles embedded within the gel---are challenging to quantify, see the measurements in \cite{jolly1996, an2012, yu2017}. Accurate material modeling of these measurements is also discussed heavily in the literature \cite{brigadnov2003, kankanala2004, saxena2014, spieler2014, sutrisno2015recent, metsch2016, schubert2016, mehnert2017, cantera2017modeling}. \\

The following free energy density is the basis of modeling materials response by using the deformation gradient, $\t F$ and the magnetic flux density, $\t B$, as follows:
\begeq \label{elastomer.mat.model}
\rho_0 \free = \frac{\mu}{4} \bigg( 1 + \tilde\alpha \tanh\Big( \frac{I_4}{\Bs} \Big) \bigg) \Big( (1+n)(I_1-3) + (1-n)(I_2-3) \Big) + q I_4 + r I_6 \ , \\
C_{ij} = F_{ji} F_{jk} \comma
I_1 = C_{ii} \comma
I_2 = \frac12 \Big( I_1^2 - C_{ij} C_{ji} \Big) \comma
I_4 = B_i B_i \comma
I_6 = C_{ij} B_j C_{ik} B_k \ .
\eqend
We have assumed an isochoric material as well as a neo-\textsc{Hooke}an mechanical response. The parameters used for composite material are
\begeq
\mu = 260\cdot10^3\,\text{Pa} \comma
\Bs = 1\,\text{T}^2 \comma
\tilde\alpha = 0.3 \comma
n = 0.3 \comma
q = r = \frac1{\mu_0} \ .
\eqend

\begin{figure} 
\centering
\includegraphics[width=0.65\linewidth]{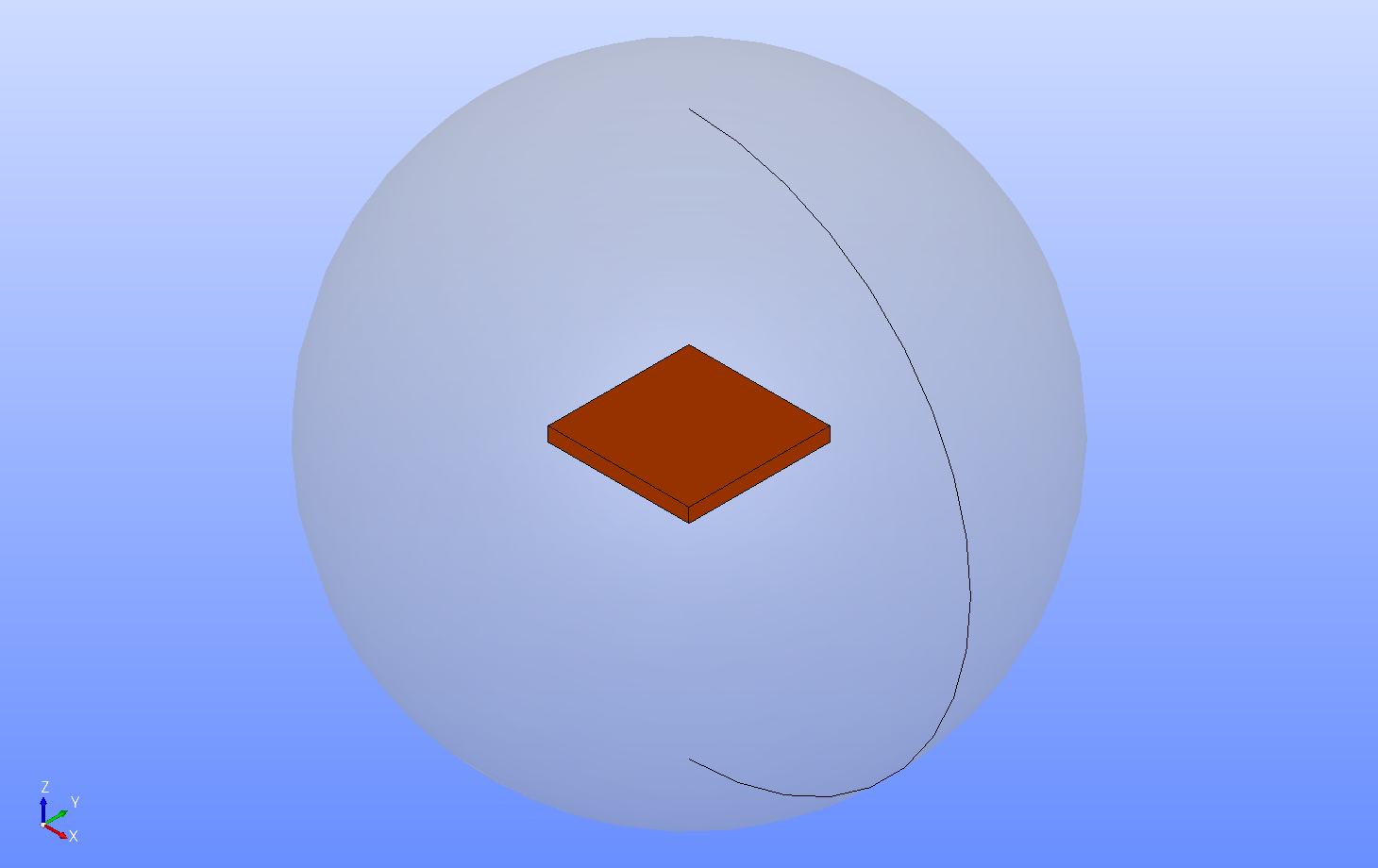}
 \caption{CAD model of the magnetorheological elastomer (orange) embedded in air (gray, transparent).}
 \label{fig:mre.1}
\end{figure}

\begin{figure} 
\centering
\includegraphics[width=0.49\linewidth]{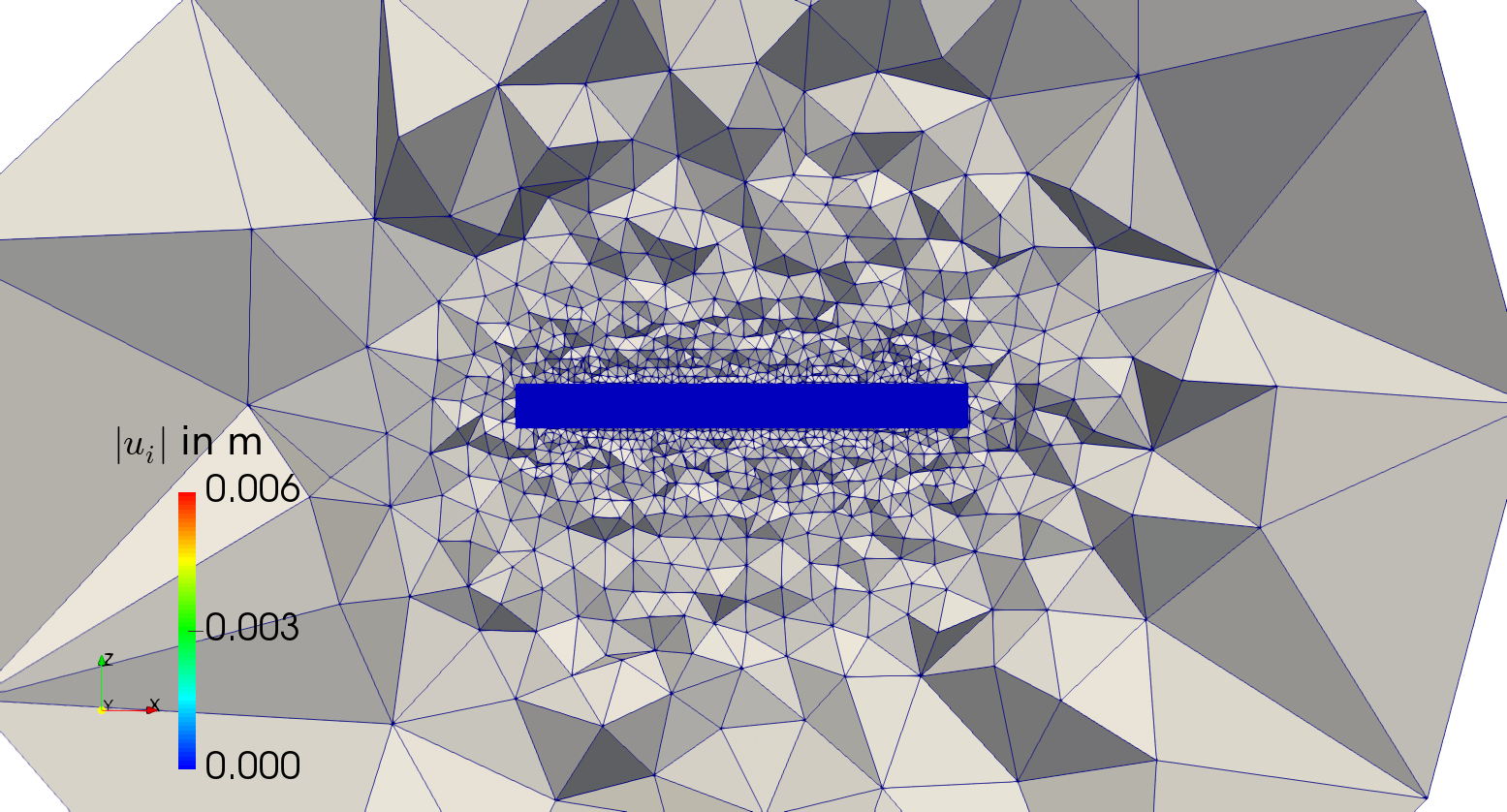}
\includegraphics[width=0.49\linewidth]{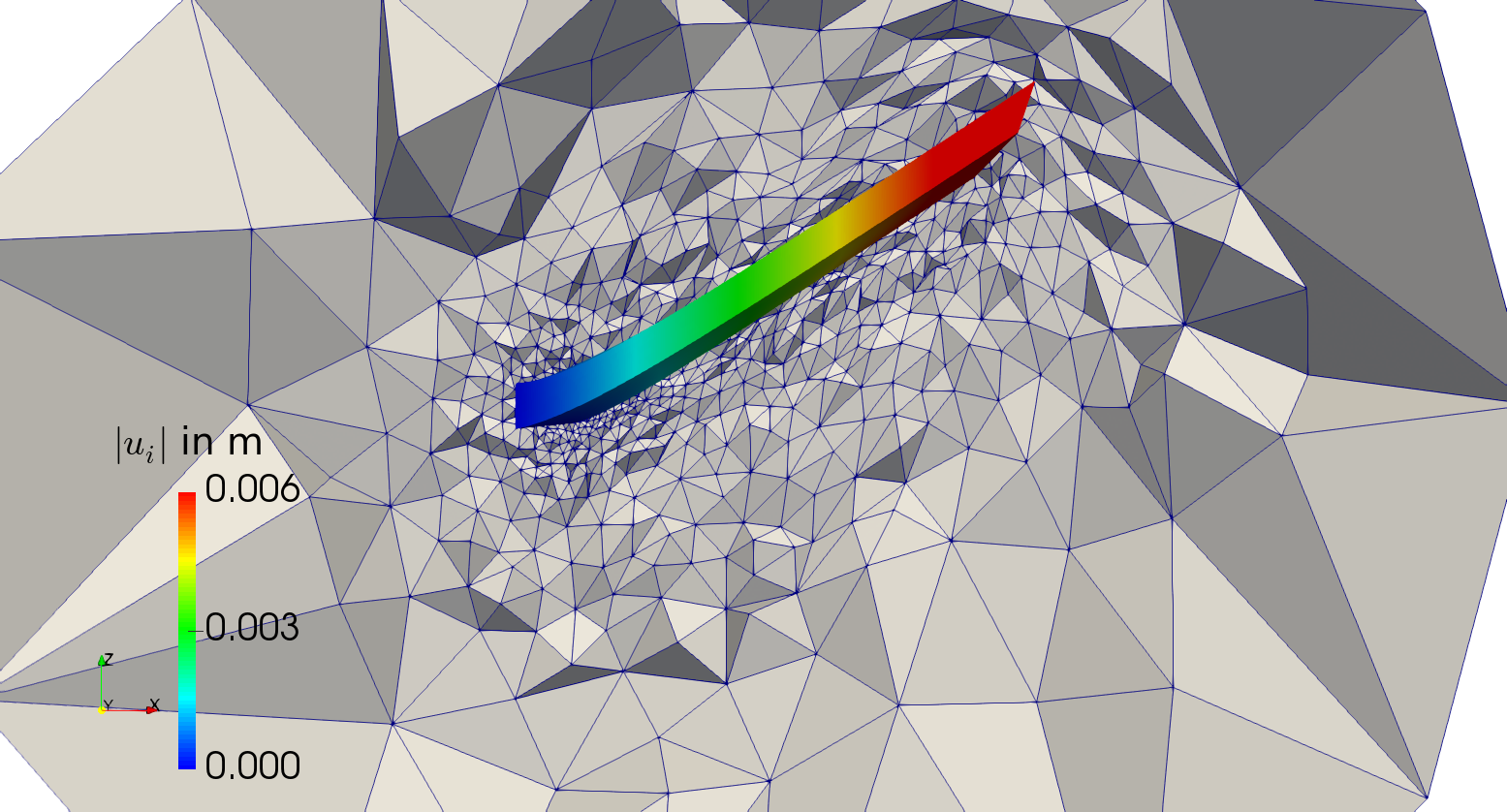}
 \caption{Configuration of the magnetorheological elastomer before and after the mechanical load is applied. The plate is colored by the magnitude of the displacement as well as morphed by the displacement {\em without} scaling. The air or vacuum mesh is not colored by any field and has a crinkle cut to present the morphing of the elements.}
 \label{fig:mre.2}
\end{figure}

\begin{figure} 
\centering
\includegraphics[width=0.49\linewidth]{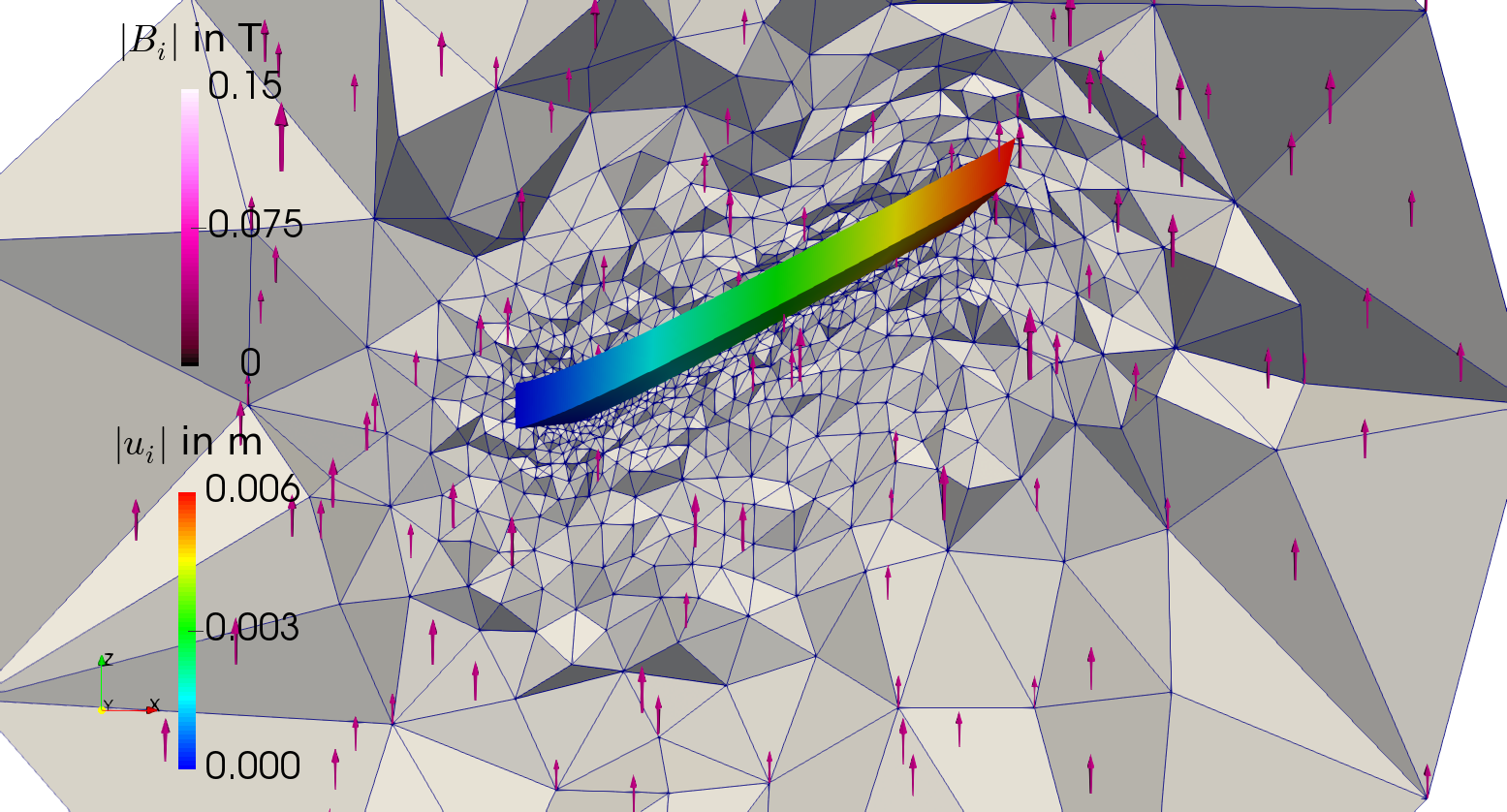}
\includegraphics[width=0.49\linewidth]{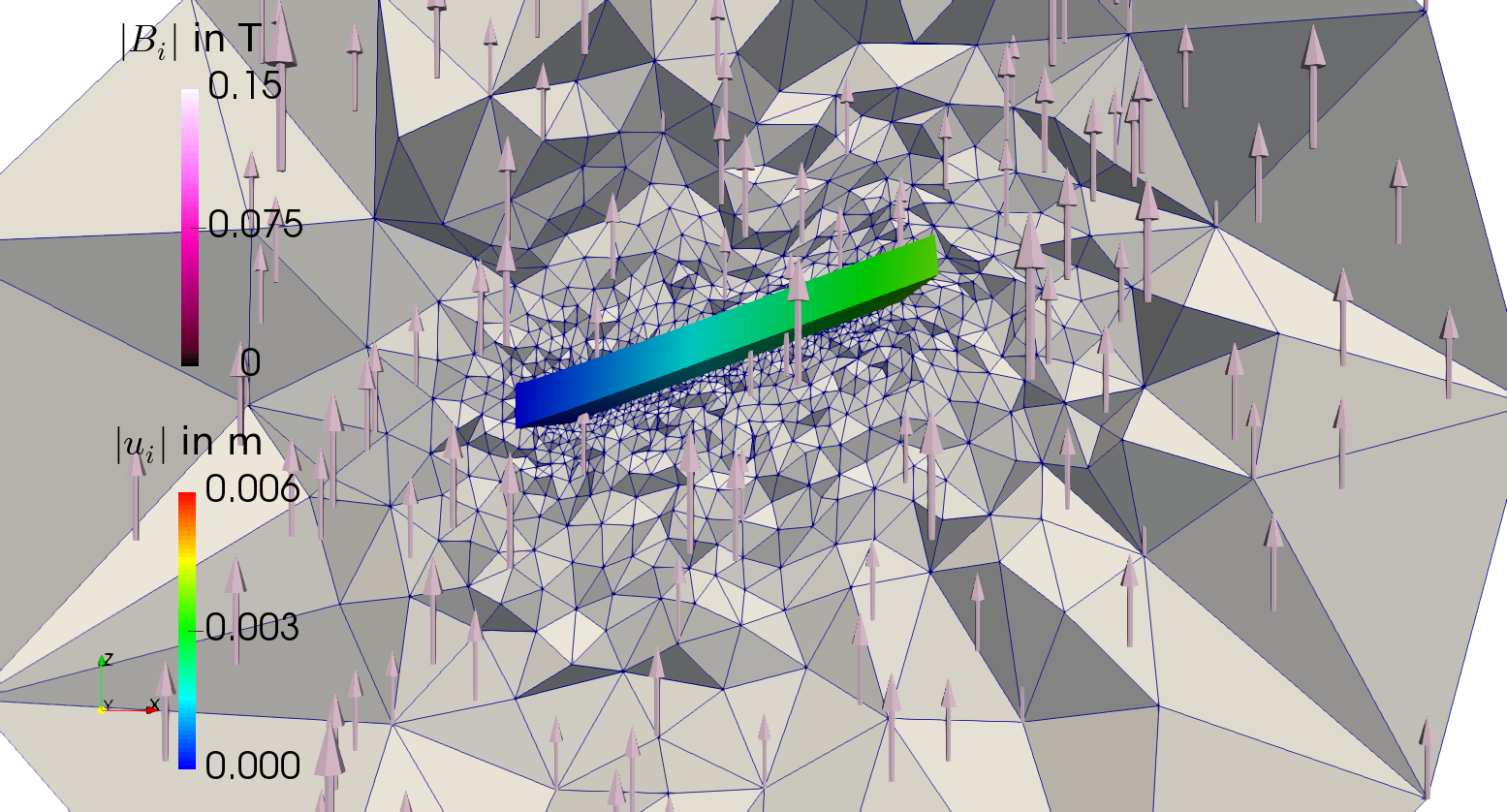}
 \caption{Configuration of the magnetorheological elastomer as the magnetic field is increased as 3\,s (left) and 6\,s (right). The arrows indicated the magnitude and orientation of the magnetic flux density $B$. The plate is colored by the magnitude of the displacement as well as morphed by the displacement {\em without} scaling. The air or vacuum mesh is not colored by any field and has a crinkle cut to present the morphing of the elements.}
 \label{fig:mre.3}
\end{figure}

A simple plate of 10\,mm$\times$10\,mm$\times$1\,mm is embedded in air as shown in Fig.\,\ref{fig:mre.1}. 
The plate is clamped on one side and a tangential traction is applied to opposite end oriented in the $z$-axis. We first apply the load with no electromagnetic fields present, deforming it from its reference state into to an initial, deformed state shown in Fig.\,\ref{fig:mre.2}. This step is performed as a nonlinear static solution of only the mechanical fields.
We emphasize that no scaling is used such that the presented deformation is the actual computed deformation. The mechanical load is held constant throughout the rest of simulation. At the outer boundaries, $\p\Om$, the following magnetic potential is applied leading to a time-varying spatially-constant magnetic flux,
\begeq
A_i = \begin{pmatrix} 0 \\ x B_\text{o} \sin(2\uppi \nu t) \\ 0 \end{pmatrix} \comma
B_i = \epsilon_{ijk} \pd{A_k}{x_j} = \begin{pmatrix} 0 \\ 0 \\ B_\text{o} \sin(2\uppi \nu t) \end{pmatrix}\comma
\forall \t x \in \p\Om \ .
\eqend
The boundary conditions are $\phi=0$ and the above form for $\t A$ using a period of 10\,s, meaning $\nu=0.1$. The deformation change at 3\,s and 6\,s is presented in Fig.\,\ref{fig:mre.3}. 
As the magnetic field increases, the body effectively stiffens leading to a smaller deformation under the same applied force. The stiffening of the structure is controlled by the material parameters in Eq.\,\eqref{elastomer.mat.model}, mainly by $\tilde\alpha$ until the saturation is achieved at $\Bs$. As seen in Fig.\,\ref{fig:mre.3}, increasing $\t B$ increases the stiffening effect, decreasing the magnitude of the deformation. 
This contactless stiffening mechanism could be used as either a sensor or actuator in a power transmission application where a winding (not included in the simulation) would be used to generate or sense the magnetic field. 

\subsection{Thermoelectric heat recovery}

\begin{figure} 
\centering
\includegraphics[width=0.49\linewidth]{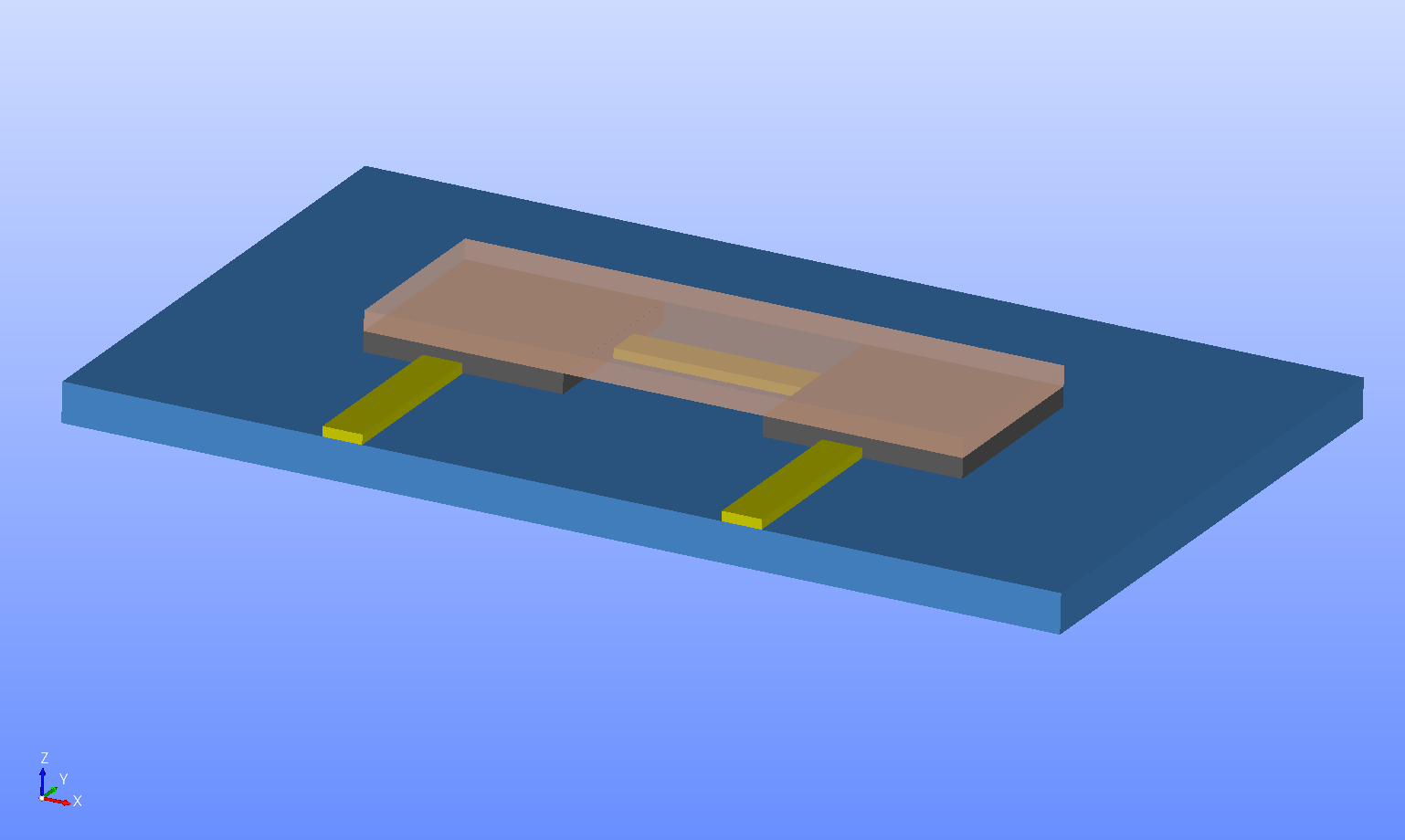}
\includegraphics[width=0.49\linewidth]{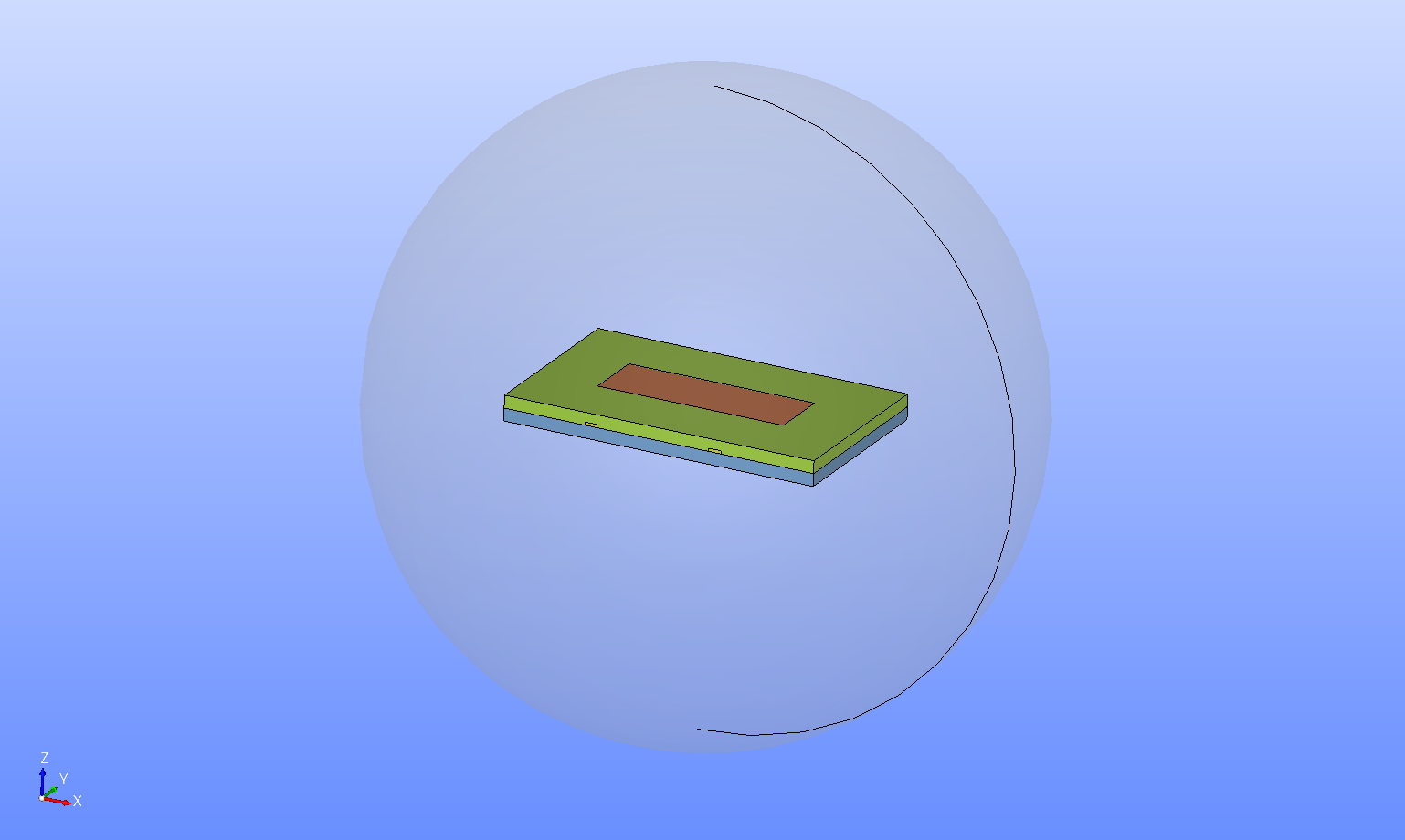}
 \caption{A simplified circuit board. Left: On the board (blue) two microchips (gray) are attached that are connected by traces (yellow). On top of the chips a thermoelectric ceramic (transparent, orange) is placed. Right: The whole assembly out of the board, chips, traces overmolded by a silicon gel (green) is embedded in air (transparent, gray).}
 \label{fig:board.1}
\end{figure}

We demonstrate the applicability of the our computational framework by simulating a thermoelectric energy recovery system suitable for use in computing servers (clusters). Especially in parallel computing, hundreds of CPUs work collectively and a significant amount of energy is dissipated from the CPUs. By using the thermoelectric effects, part of the dissipating energy can be recovered. The device is an assembly of two integrated circuits joined by copper traces, and a thermoelectric ceramic mounted on top of a substrate. The whole assembly is over-molded in a silicon gel in order to reduce the environmental effects like corrosion. The device is shown in Fig\,\ref{fig:board.1}, left is the electronic part and right is the mold cutting the contact of electronic parts with the environment. An alternating current (AC) is used with an electric potential difference of 12\,V on the trace endings in front as seen in Fig.\,\ref{fig:board.1}. During operation, \textsc{Joule} heating causes a temperature increase on the microchip leading to an electric current across the piezoceramic sheet measured as a potential difference. This sheet includes an assembly of conductive materials with a thermoelectric constant $\pi$ generating an electric current in Eq.\,\eqref{current.polarization} because of a temperature difference. Even in this very simple model, there are different materials and several interfaces. The board is a composite material, mostly it is made of glass reinforced epoxy resin. The microchips are represented as ceramic materials without the detailed internal assembly. Copper traces are used. \\

\begin{figure} 
\centering
\includegraphics[width=0.7\linewidth]{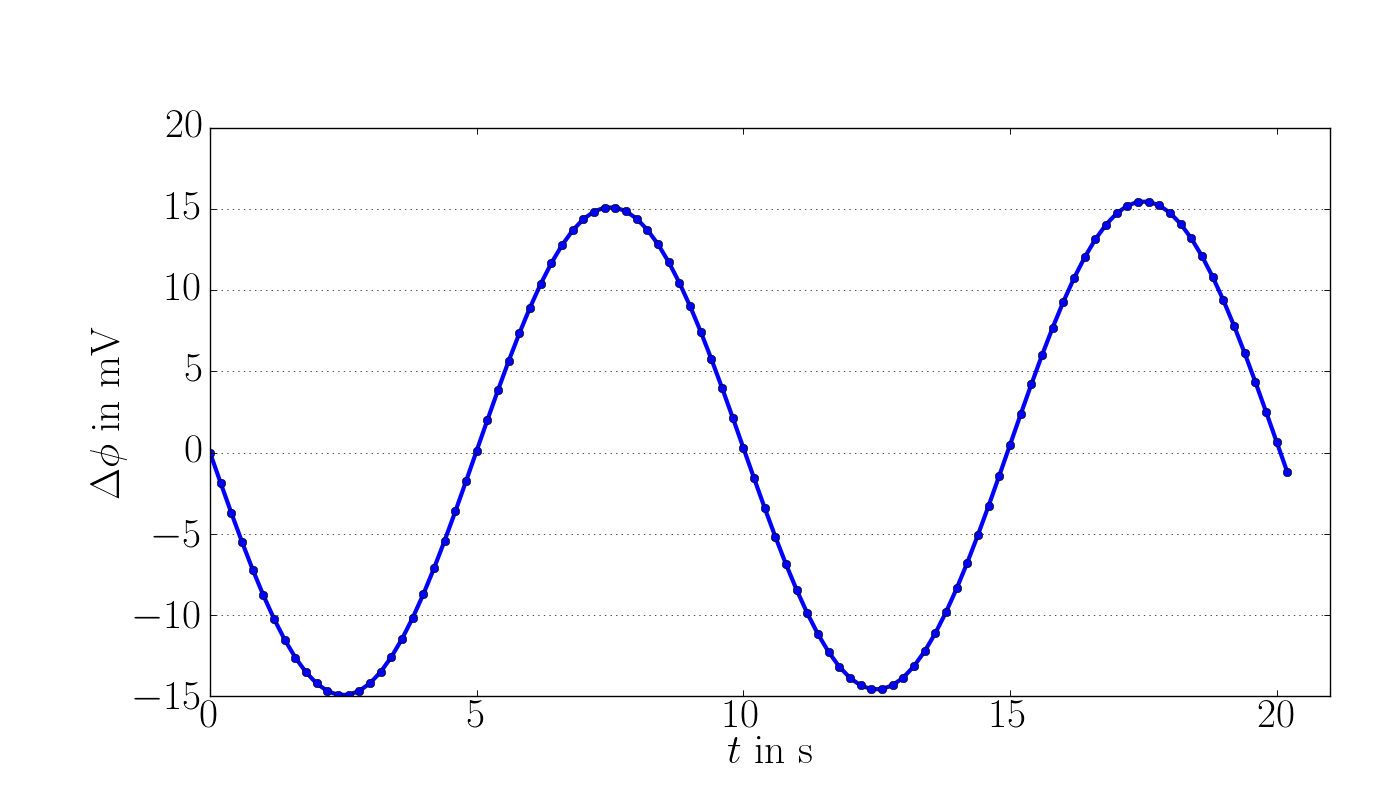}
 \caption{Output of the thermoelectric device, electric potential difference, $\Delta \phi$ in mV, as a result of AC put over the traces.}
 \label{fig:board.2}
\end{figure}

An electric potential difference is imposed at the endpoints of traces in front side of the device (apparent in Fig.\,\ref{fig:board.1} (left)) by using \textsc{Dirichlet} boundary conditions. This difference creates an electric current through the microchips leading to \textsc{Joule}'s heating. As a result, the microchips heat up the thermoelectric ceramic sheet from below, at the same time, on top of the sheet a cooling agent (like water in a closed system) reduces the temperature by a mixed boundary condition, $h(T-T_\Ref)$, simulated with a high convection constant $h=10^5$\,J/(s\,m$^2$\,K). The temperature difference over the thermoelectric sheet generates a heat flux and an electric current, which results in an electric potential difference across the top and bottom layers of the thermoelectric sheet. We record this induced difference over time and present in Fig.\,\ref{fig:board.2}.
The thermoelectric energy harvesting shown in this simulation is not very effective; however, it can be used in clusters just to recover a small fraction of the dissipating energy. We stress that this coupling is inherent in the material and no degradation is expected to occur. Hence, it is potentially of interest to use the initial investment to enable energy recovery over the total service life of the cluster.

\section{Conclusion}
\label{sec:conclusion}
We have developed a complete theory of continuum mechanics with electromagnetic interaction in solid bodies under large deformations. Balance equations for mechanical, thermal, and electromagnetic fields have been discussed and all necessary constitutive equations have been derived by exploiting thermodynamical principles. The constitutive equations are general and involve all coupling phenomena resulting in piezoelectric, pyroelectric, and thermoelectric responses. Some irreversible effects are neglected, such as plasticity, viscoelasticity, and material hysteresis with respect to the electromagnetic fields. The proposed approach is general for polarized elastic materials subject to large deformations. The formulation results in a nonlinear variational form with complete coupling of all of the mechanical and electromagnetic fields. In order to reduce the computational cost, we have proposed a novel method and solve electromagnetic fields in the whole domain, whereas the deformation and temperature are solved only within the continuum body. Separation of field and matter is possible; however, their equations are defined in different placements. Therefore, from the very beginning of the development of the theory, we have emphasized different frames by using an explicit notation. Numerical solution is possible in the \textsc{Lagrange}an and \textsc{Euler}ian frames by using a staggered scheme with an appropriate mesh morphing algorithm. The nonlinear variational forms have been solved for thermomechanical and electromagnetic fields by using the finite element method with the aid of open-source packages developed under the FEniCS project. In order to encourage further achievements, the mesh morphing algorithm implementation is released under the GNU LGPL \url{https://github.com/afqueiruga/afqsfenicsutil} and the scripts and geometry files that performed the simulations are released under the GNU GPL at the repository \url{https://github.com/afqueiruga/EMSI-2018}.

\section*{Acknowledgement}
B. E. Abali's work was partly supported by a grant from the Daimler and Benz Foundation.

\section*{References}
\bibliographystyle{plainnat}
\bibliography{Abali_EM}

\appendix

\section{Electric current due to the bound charges} \label{app.el.current.def}

By using
\begeq
D_i = \mD_i - P_i \comma H_i = \mH_i + \MM_i 
\eqend
in \textsc{Maxwell}'s equations 
\begeq
\pd{\mD_i}{x_i} - \pd{P_i}{x_i} = \rho z^\fr + \rho z^\bo \ , \\
-\pd{\mD_i}{t} + \pd{P_i}{t} + \epsilon_{ijk} \pd{\mH_k}{x_j} + \epsilon_{ijk} \pd{\MM_k}{x_j} = J_i \ ,
\eqend
we realize
\begeq
\pd{\mD_i}{x_i} = \rho z^\fr \comma - \pd{P_i}{x_i} = \rho z^\bo \ , \\
-\pd{\mD_i}{t} + \epsilon_{ijk} \pd{\mH_k}{x_j}  = J_i - \pd{P_i}{t} -  \epsilon_{ijk} \pd{\MM_k}{x_j} = J_i^\fr \ .
\eqend

\section{Balance of electromagnetic momentum} \label{app.elmo.balance}

We start by obtaining the time rate of the chosen electromagnetic momentum 
\begeq
\pd{\elmo_i}{t} = \pd{\epsilon_{ijk} \mD_j B_k }{t} 
= \epsilon_{ijk} \pd{D_jB_k}{t}  + \epsilon_{ijk} \pd{P_jB_k}{t} \ .
\eqend
The first term can be rewritten, by using 
\begeq
\epsilon_{ijk}=-\epsilon_{ikj} \comma
\epsilon_{ijk}\epsilon_{klm} = \delta_{il}\delta_{jm} - \delta_{im}\delta_{jl} \ ,
\eqend
as well as \textsc{Maxwell}'s equations
\begeq
\epsilon_{ijk} \pd{D_jB_k}{t} 
= \epsilon_{ijk} \Big( \epsilon_{jlm} \pd{H_m}{x_l} - J_j \Big) B_k - \epsilon_{ijk} D_j \epsilon_{klm} \pd{E_m}{x_l}
= \\
= -\pd{H_k}{x_i} B_k + \pd{H_i}{x_k} B_k - (\t J\times \t B)_i - D_j \pd{E_j}{x_i} + D_j \pd{E_i}{x_j} \ .
\eqend
Moreover, by using \textsc{Maxwell--Lorentz} aether relations we observe
\begeq
\pd{H_k}{x_i} B_k = \frac1{\mu_0} \pd{B_k}{x_i} B_k =\frac1{\mu_0} \pd{}{x_i} \Big(\frac12 B_k B_k\Big) = \frac12 \pd{H_k B_k}{x_i} \ , \\
D_j \pd{E_j}{x_i} = \eps_0 E_j \pd{E_j}{x_i} = \eps_0 \pd{}{x_i} \Big( \frac12 E_j E_j\Big) = \frac12 \pd{D_j E_j}{x_i} \ .
\eqend
Finally, by utilizing \textsc{Maxwell}'s equations we achieve
\begeq
\pd{H_i}{x_k} B_k = \pd{H_i B_k}{x_k} \ , \\
D_j \pd{E_i}{x_j} = \pd{D_j E_i}{x_j} - \rho z E_i \ .
\eqend
By combining all above, we obtain
\begeq
\pd{\elmo_i}{t}
= \pd{}{x_j}\underbrace{\Big( -\frac12 \delta_{ij} \big( D_k E_k + H_k B_k \big) + H_i B_j + D_j E_i \Big)}_{m_{ji}}
-\underbrace{\Big( \rho z E_i + (\t J\times \t B)_i  - \pd{(\t P\times \t B)_i}{t} \Big)}_{\elF_i} \ ,
\eqend
after comparing to the the balance of electromagnetic momentum.

\section{Balance of electromagnetic energy}\label{app.elen.balance}

By starting with the divergence of the chosen electromagnetic flux
\begeq
\pd{\Po_i}{x_i} = \epsilon_{ijk} \pd{\mH_j E_k}{x_i} = \epsilon_{ijk} \pd{H_j E_k}{x_i} - \underbrace{\epsilon_{ijk} \pd{\MM_j E_k}{x_i} }_{ \displaystyle \pd{(\tt\MM\times \t E)_i}{x_i}} \ , 
\eqend
we can rewrite the first term by using $\epsilon_{ijk}=\epsilon_{jki}=\epsilon_{kij}$ and $\epsilon_{ijk}=-\epsilon_{jik}$, as well as inserting \textsc{Maxwell}'s equations,
\begeq
\epsilon_{ijk} \pd{H_j E_k}{x_i} 
= \epsilon_{kij} \pd{H_j}{x_i} E_k - H_j \epsilon_{jik}  \pd{E_k}{x_i}
= \Big( \pd{D_k}{t} + J_k \Big) E_k + H_j \pd{B_j}{t}
= \\
= \pd{}{t} \Big( \frac12 D_k E_k + \frac12 H_j B_j \Big) + J_k E_k \ .
\eqend
Since we have
\begeq
J_i E_i = \Big( J_i^\fr + \pd{P_i}{t} + \epsilon_{ijk} \pd{\MM_k}{x_j} \Big) E_i 
= J_i^\fr E_i + \pd{P_i E_i}{t} - P_i \pd{E_i}{t} + \epsilon_{ijk} \pd{\MM_k E_i}{x_j} - \MM_k \epsilon_{ijk} \pd{E_i}{x_j}  
= \\
= J_i^\fr E_i + \pd{P_i E_i}{t} - P_i \pd{E_i}{t} + \epsilon_{jki} \pd{\MM_k E_i}{x_j} + \MM_k \epsilon_{kji} \pd{E_i}{x_j}  
= \\
= J_i^\fr E_i + \pd{P_i E_i}{t} - P_i \pd{E_i}{t} + \pd{(\tt\MM\times\t E)_j}{x_j} - \MM_k \pd{B_k}{t}
= \\
= J_i^\fr E_i + \pd{}{t}(P_i E_i-\MM_i B_i) - P_i \pd{E_i}{t} + \pd{(\tt\MM\times\t E)_j}{x_j} + B_k \pd{\MM_k}{t} \ ,
\eqend
we obtain
\begeq
\pd{\Po_i}{x_i} = \pd{}{t} \Big( P_i E_i-\MM_i B_i + \frac12 (D_k E_k + H_j B_j) \Big) + J_i^\fr E_i - P_i \pd{E_i}{t} + B_k \pd{\MM_k}{t} \ , \\
\pd{}{t} \underbrace{\Big( P_i E_i-\MM_i B_i + \frac12 (D_k E_k + H_j B_j) \Big)}_{\displaystyle e^\field} = \pd{\Po_i}{x_i} - \underbrace{\Big( J_i^\fr E_i - P_i \pd{E_i}{t} + B_k \pd{\MM_k}{t} \Big)}_{\displaystyle\pi}\ ,
\eqend

\section{Maxwell symmetry} \label{maxwell.symmetry}
The \textsc{Maxwell} symmetry relations can be seen by using simple relations,
\begeq
\tilde c^{21}_{ji} = \pd{n_{ji}}{T} = \pd{^2 \free}{T \p F_{ij}} =  \pd{^2 \free}{F_{ij} \p T} = - \pd{\eta}{F_{ij}} = -\tilde c^{12}_{ij} 
\ , \\
\tilde c^{31}_{i} = \pd{p_{i}}{T} = -\pd{^2 \free}{T \p E_{i}} =  -\pd{^2 \free}{E_{i} \p T} = \pd{\eta}{E_{i}} = \tilde c^{13}_{i} 
\ , \\
\tilde c^{41}_{i} = \pd{\mm_{i}}{T} = -\pd{^2 \free}{T \p B_{i}} =  -\pd{^2 \free}{B_{i} \p T} = \pd{\eta}{B_{i}} = \tilde c^{14}_{i} 
\ , \\
\tilde c^{51} = \pd{\pre}{T} = -\pd{^2 \free}{T \p \v} =  -\pd{^2 \free}{\v \p T} =  \pd{\eta}{\v} = \tilde c^{15} 
\ , \\
\eqend
furthermore,
\begeq
\tilde c^{32}_{ikl} = \pd{p_{i}}{F_{kl}} = -\pd{^2 \free}{F_{kl} \p E_{i}} =  -\pd{^2 \free}{E_i \p F_{kl}} = -\pd{n_{lk}}{E_i} = -\tilde c^{23}_{lki} 
\ , \\
\tilde c^{42}_{ikl} = \pd{\mm_{i}}{F_{kl}} = -\pd{^2 \free}{F_{kl} \p B_{i}} =  -\pd{^2 \free}{B_i \p F_{kl}} = -\pd{n_{lk}}{B_i} = -\tilde c^{24}_{lki} 
\ , \\
\tilde c^{52}_{kl} = \pd{\pre}{F_{kl}} = -\pd{^2 \free}{F_{kl} \p \v} =  -\pd{^2 \free}{\v \p F_{kl}} = -\pd{n_{lk}}{\v} = -\tilde c^{25}_{lk} 
\ , 
\eqend
as well as
\begeq
\tilde c^{43}_{ik} = \pd{\mm_{i}}{E_{k}} = -\pd{^2 \free}{E_{k} \p B_{i}} =  -\pd{^2 \free}{B_i \p E_{k}} = \pd{p_{k}}{B_i} = \tilde c^{34}_{ki} 
\ , \\
\tilde c^{53}_{k} = \pd{\pre}{E_{k}} = -\pd{^2 \free}{E_{k} \p \v} =  -\pd{^2 \free}{\v \p E_{k}} = \pd{p_{k}}{\v} = \tilde c^{35}_{k} 
\ , 
\eqend
and
\begeq
\tilde c^{54}_{k} = \pd{\pre}{B_{k}} = -\pd{^2 \free}{B_{k} \p \v} =  -\pd{^2 \free}{\v \p B_{k}} = \pd{\mm_{k}}{\v} = \tilde c^{45}_{k} 
\ .
\eqend

\end{document}